\documentclass[onecolumn]{IEEEtran}
\usepackage{rotating}
\usepackage [mathscr] {eucal}
\usepackage{mathpazo}
\usepackage{times}
\usepackage{amsmath}
\usepackage{amsfonts}
\usepackage{latexsym}
\usepackage{amssymb}
\usepackage[noadjust]{cite}
\usepackage[ruled,vlined]{algorithm2e}
\usepackage{upref}
\usepackage{theorem}
\usepackage{graphicx}
\usepackage{multirow}
\usepackage{overpic}
\usepackage[usenames]{color}
 \usepackage {tikz}
\usetikzlibrary {positioning}
\usepackage{mathtools}
\usepackage{xparse}
\usepackage{optidef}
\DeclarePairedDelimiterX{\set}[1]{\{}{\}}{\setargs{#1}}
\NewDocumentCommand{\setargs}{>{\SplitArgument{1}{;}}m}
{\setargsaux#1}
\NewDocumentCommand{\setargsaux}{mm}
{\IfNoValueTF{#2}{#1} {#1\,\delimsize|\,\mathopen{}#2}}

\DeclarePairedDelimiter\abs{\lvert}{\rvert}
\DeclarePairedDelimiter\ceil{\lceil}{\rceil}
\DeclarePairedDelimiter\floor{\lfloor}{\rfloor}
\DeclarePairedDelimiter\parenv{\lparen}{\rparen}
\DeclarePairedDelimiter\sparenv{\lbrack}{\rbrack}

\theoremstyle{plain}
\newtheorem{theorem}{Theorem}
\newtheorem{lemma}[theorem]{Lemma}
\newtheorem{prop}[theorem]{Proposition}

\newtheorem{definition}[theorem]{Definition}
\newtheorem{example}{Example}
\newtheorem{remark}{Remark}

\newcommand{\cE}{\mathcal{E}}
\newcommand{\cF}{\mathcal{F}}

\newcommand{\cP}{\mathcal{P}}

\renewcommand{\le}{\leqslant}
\renewcommand{\leq}{\leqslant}
\renewcommand{\ge}{\geqslant}
\renewcommand{\geq}{\geqslant}

\newcommand{\N}{\mathbb{N}}
\newcommand{\R}{\mathbb{R}}
\newcommand{\Z}{\mathbb{Z}}
\newcommand{\Pro}{\mathbb{P}}

\newcommand{\src}{\sigma}
\newcommand{\tar}{\tau}
\DeclareMathOperator{\MB}{MB}

\DeclareMathOperator{\wt}{wt}

\newcommand{\E}{\mathbb{E}}
\newcommand{\eqdef}{\triangleq}

\newcommand{\oee}{\overline{e}}

\newcommand{\oc}{\overline{c}}

\newcommand{\bx}{\mathbf{x}}
\newcommand{\by}{\mathbf{y}}
\newcommand{\bX}{\mathbf{X}}
\newcommand{\bzero}{\mathbf{0}}
\newcommand{\bY}{\mathbf{Y}}
\newcommand{\ov}{\overline{v}}

\newcommand{\ou}{\overline{u}}

\newcommand{\ow}{\overline{w}}

\newcommand{\ox}{\overline{x}}

\newcommand{\oY}{\overline{Y}}
\newcommand{\oy}{\overline{y}}

\newcommand{\ozero}{\overline{0}}

\newcommand{\sB}{\mathscr{B}}
\newcommand{\rep}{\mathrm{rep}}
\newcommand{\muni}{\mu^{\mathsf{u}}}

\begin{document}


\title{Quantized-Constraint Concatenation \\ and the Covering Radius of Constrained Systems}

\author{
  Dor Elimelech~\IEEEmembership{Student Member,~IEEE}, Tom Meyerovitch, and Moshe Schwartz~\IEEEmembership{Senior Member,~IEEE}%
  \thanks{Dor Elimelech is with the School
    of Electrical and Computer Engineering, Ben-Gurion University of the Negev,
    Beer Sheva 8410501, Israel
    (e-mail: doreli@post.bgu.ac.il).}%
  \thanks{Tom Meyerovitch is with the  Department of Mathematics, 
    Ben-Gurion University of the Negev,
    Beer Sheva 8410501, Israel
    (e-mail: mtom@bgu.ac.il).}%
  \thanks{Moshe Schwartz is with the School
    of Electrical and Computer Engineering, Ben-Gurion University of the Negev,
    Beer Sheva 8410501, Israel
    (e-mail: schwartz@ee.bgu.ac.il).}%
  \thanks{The paper was submitted in part to ISIT 2023.}
}

\maketitle

\begin{abstract}
We introduce a novel framework for implementing error-correction in constrained systems. The main idea of our scheme, called Quantized-Constraint Concatenation (QCC), is to employ a process of embedding the codewords of an error-correcting code in a constrained system as a (noisy, irreversible) quantization process. This is in contrast to traditional methods, such as concatenation and reverse concatenation, where the encoding into the constrained system is reversible. The possible number of channel errors QCC is capable of correcting is linear in the block length $n$, improving upon the $O(\sqrt{n})$ possible with the state-of-the-art known schemes. For a given constrained system, the performance of QCC depends on a new fundamental parameter of the constrained system -- its covering radius. 

Motivated by QCC, we study the covering radius of constrained systems in both combinatorial and probabilistic settings. We reveal an intriguing characterization of the covering radius of a constrained system using ergodic theory. We use this equivalent characterization in order to establish efficiently computable upper bounds on the covering radius. 
\end{abstract}



\section{Introduction}
\label{sec:intro}
\IEEEPARstart{C}{onstrained} codes are often employed in communication and storage systems in  order to mitigate the occurrence of data-dependent errors. In many channels, some words are more prone to error than others, and therefore by avoiding them, the number of errors is reduced. Such codes are called constrained codes.  While the use of constrained codes may significantly reduce the occurrence of data-dependent errors, in many realistic scenarios, the transmitted data may still be corrupted by data-independent errors.

A well-known strategy for handling the corruption of data is to combine error-correcting codes with constrained codes. This has been extensively studied during the past 40 years (see for example \cite{bliss1981circuitry, mansuripur1991enumerative, immink1997practical, fan1998modified,chee2020efficient,gabrys2020segmented}), and recently regained attention due to the increased interest in DNA storage systems. Over the last years, error-correcting constrained codes for DNA storage have been studied in numerous works \cite{cai2021correcting, lu2021design, benerjee2022homopolymers, li2022multiple, cai2021coding, nguyen2021capacity, nguyen2020constrained,weber2020single,press2020hedges }, with 
particular attention given to the GC-content constraint and the run-length (homopolymer) constraint.

Despite the considerable recent progress made in construction and analysis of error-correcting constrained codes for specific families of constraints, still, only a few general frameworks for implementing error correction in constrained systems are known (see~\cite[Ch.~8]{marcus2001introduction} for a survey). An important example for such a framework is the method of reverse concatenation, sometimes called modified concatenation (see \cite{bliss1981circuitry, mansuripur1991enumerative, immink1997practical, fan1998modified}), in which an error-correction encoding follows a constrained encoder. Recently, an improvement of the reverse-concatenation method, called segmented reverse concatenation, was suggested \cite{gabrys2020segmented}. A principle limitation to these methods is in the error-correction capability. While the 
state-of-the-art method presented in \cite{gabrys2020segmented} allows for a correction of $O(\sqrt{n})$ errors (where $n$ is the block length), a general technique for correcting $\Theta(n)$ errors in constrained systems is unknown.   

Motivated by this gap, we propose an alternative strategy, \emph{quantized-constraint concatenation (QCC)}, for the implementation of error correction in constrained systems, which also works in the presence of $\Theta(n)$ errors. The basic idea behind our proposed method is simple: we suggest to consider the embedding process of information in the constrained media as a \emph{quantization} process, rather than a coding process. In traditional methods (including concatenation and reverse concatenation) a constrained word represents the data to be transmitted and protected against errors. Thus, the constrained encoder is reversible, and it incurs a rate penalty, on top of the rate penalty for the error-correcting code. In QCC, we consider the constrained word as a corrupted version of the information, obtained by a quantization procedure. Thus, the constrained quantizer incurs no rate penalty. Instead, the parameters of the error-correcting code are designed to handle both errors caused by the channel and by the quantization process.

Let $\sB_n\subseteq \Sigma^n$ be some set of constrained words of length $n$ over some finite alphabet $\Sigma$. Assume furthermore that  $r<n$ is an integer such that for any word $\oy\in \Sigma^n$ there exists a corresponding word $\ox\in \sB_n$ with Hamming distance $d(\ox,\oy)\leq r$. Given an error-correcting code $C\subseteq \Sigma^n$ that can correct $t>r$ errors, we propose the following constrained error-correction procedure (see Figure~\ref{fig:Block}): 

\begin{figure}[t]
\centering
\begin{overpic}[scale=0.3]
    {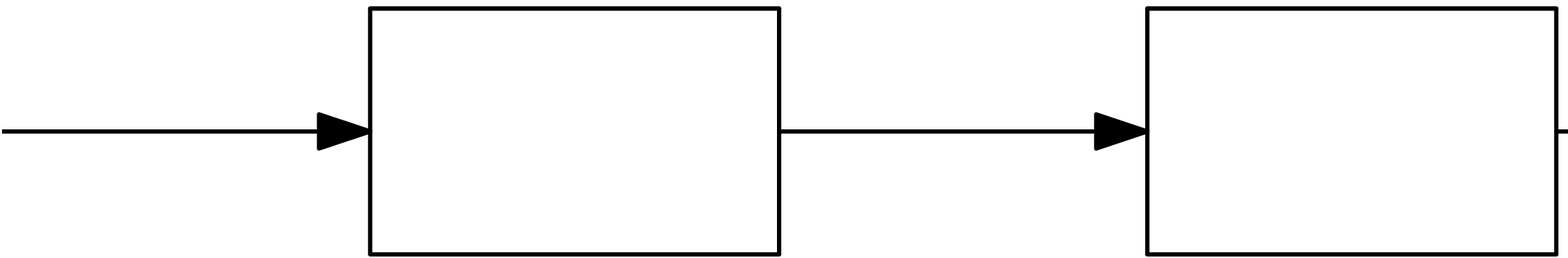}
    \put(4,4.5){$\ow$}
    \put(12.3,2.7){$\begin{matrix} \text{ECC }\\\text{Encoder }\end{matrix}$}
    \put(24.2,4.5){$\oy\in C$}
    \put(34,3.2){Quantizer}
    \put(46,4.5){$\ox\in \sB_n$}
    \put(57,3.2){Channel}
    \put(71,4.5){$\ox'$}
    \put(79.2,2.7){$\begin{matrix} \text{ECC }\\\text{Decoder }\end{matrix}$}
    \put(93,4.5){$\ow'$}
\end{overpic}
\caption{
A block diagram describing our proposed error correction procedure.}
\label{fig:Block}
\end{figure}

\begin{itemize}
    \item \emph{Encoding:} Given an information  word $\ow$, use an encoder for an error-correcting code to map it to a codeword $\oy\in C$.
    \item \emph{Quantization:} Given $\oy\in C$, find a constrained word $\ox\in \sB_n$ such that $d(\oy,\ox)\leq r$, and transmit $\ox$. 
    \item \emph{Channel:} At the channel output, $\ox'\in\Sigma^n$, a corrupted version of $\ox$, is observed.
    \item \emph{Decoding:} Use the decoder for $C$ on $\ox'$ and obtain $\ow'$.
\end{itemize}

If the channel does not introduce more than $t-r$ errors, i.e., $d(\ox,\ox')\leq  t-r$, then $d(\oy,\ox')\leq t$. Since $C$ can correct $t$ errors, we have $\ow=\ow'$, namely, it is possible to correct $t-r$ channel errors. We are therefore interested in the minimal number $r$, such that any word in the space can be quantized to a word in $\sB_n$ with at most $r$ coordinates changed. In coding-theory terminology, this quantity, denoted by $R(\sB_n)$, is called the \emph{covering radius} of $\sB_n$. Using this technique, it is now possible to correct $\Theta(n)$ errors: assume that we have a constrained system such that for all $n$ we have $R(\sB_n)\leq \rho\cdot n$, and $(C_n)_{n\in\N}$ is a sequence of codes capable of correcting $\delta\cdot n$ errors for some $\delta>\rho$. Using the scheme presented above, it is therefore possible to correct $(\delta-\rho)\cdot n$ channel errors, which is linear in $n$. 

In order to further our understanding of the proposed scheme, it is crucial for us to study the covering radius of constrained systems, which is the goal of this paper. We outline the contributions we make. In Section~\ref{sec:CovRad}, we provide a combinatorial definition for the covering radius of a constrained system, and investigate some of its fundamental properties. We also observe an intriguing phenomenon: We present an example of a constrained system with positive capacity that has the same covering radius has the repetition code, which has zero capacity.

Inspired by this phenomenon, in Section~\ref{sec:AverageCov} we take a probabilistic approach and define the \emph{essential} covering radius. We show that this version disregards the extreme cases causing the unwanted phenomenon described above. We use the framework of ergodic theory to give an alternative characterization of the essential covering radius. In Section~\ref{sec:CoupAlg} we use our ergodic-theoretic definition of the essential covering radius establish upper bounds on the essential covering radius in common scenarios. Using a Markov-chain approach, we derive a general upper bound which is efficiently computable as a solution of a linear program. We also provide bounds using sliding-block-code functions. We show that in the primitive case, these bounds asymptotically attain the essential covering radius.


\section{Preliminaries}\label{sec:prem}

Throughout this paper we shall use lower-case letters, $x$, to denote scalars and symbols, overlined lower-case letters, $\ox$, to denote finite-length words, and  bold lower-case letters, $\bx$, to denote bi-infinite sequences. We use upper-case letters, $X$, for constrained systems. For a bi-infinite sequence $\bx=\dots,\bx_{-1},\bx_0,\bx_1,\dots$ and $n\leq m$ we denote the subword $\bx_{n}^m\eqdef \bx_n,\dots,\bx_m$ (and similarly $\ox_{n}^m$ for finite words). We use $\Sigma$ to denote a finite alphabet, and $[n]\eqdef \set*{0,1,\dots,n-1}$. 

The set of words of length $n$ over $\Sigma$ is denoted by $\Sigma^n$. If $\ou\in\Sigma^n$, we shall index its letters by $[n]$, i.e., $\ou=\ou_0,\ou_1,\dots,\ou_{n-1}$. For any $\ov,\ou\in \Sigma^n$, we define the Hamming distance as
\[d(\ou,\ov)\eqdef \abs*{\set*{ i\in[n] ; \ou_i\neq \ov_i}}.\]
The ball of radius $r$ (with respect to the Hamming distance) centered in $\ox$ is denoted by $B_r(\ox)$. The covering radius of a code $C\subseteq \Sigma^n$ is the minimal integer $r$ such that the union of balls of radius $r$, centered at the codewords of $C$, covers the whole space. That is,
\[R(C)\eqdef \min\set*{r\in \N\cup\set{0}; \bigcup_{\oc\in C} B_r(\oc)=\Sigma^n}.\]
Elements in $\Sigma^n$ whose distance to the closest codeword of $C$ is $R(C)$, are often  called \emph{deep holes} (e.g., see~\cite[Definition 2.1.3]{Cohen}).

We turn to discuss constrained systems. These are often studied in the framework of symbolic dynamics (see for example \cite{lind2021introduction, marcus2001introduction}). In a typical (one dimensional) setting we have a finite alphabet $\Sigma$, and the space of bi-infinite sequences of $\Sigma$, denoted $\Sigma^{\Z}$, is considered as a compact metrizable topological space, equipped with the product topology (where $\Sigma$ has the discrete topology). The dynamics on the system $\Sigma^\Z$ are realized by the shift transformation, 
  $T:\Sigma^\Z\to \Sigma^\Z$, defined by 
\[ (T\bx)_n\eqdef \bx_{n+1},  \]
which is a topological homeomorphism of the system. For a finite word $\ox\in \Sigma^{n}$ we let $[\ox]$ denote the cylinder set defined by $\ox$, which is
\begin{equation}
\label{def:cylinder}
[\ox]\eqdef \set*{\bx\in \Sigma^\Z ; \bx_{0}^{n-1}=\ox}.
\end{equation}

A subshift (or shift space) $X\subseteq \Sigma^\Z$ is a compact subspace, which is invariant under the shift transformation. For a subshift $X$, the language of $X$ is the set of all finite words that appear as subwords of some element in $X$. That is \[\sB(X)\eqdef \set*{\ox=(x_0\dots x_k) ; \exists \bx\in X, n\in \Z \text{ such that } \bx_n^{n+k}=\ox, k\in \N\cup\set{0}}.\]
The set of words of length $n$ in the language is denoted by  $\sB_n(X)\eqdef \sB(X)\cap \Sigma^n$. The topological entropy, also called capacity, of $X$ is defined to be the following limit (which exists by Fekete's lemma)
\[h(X)\eqdef\lim_{n\to\infty}\frac{\log_{\abs{\Sigma}}|\sB_n(X)|}{n}.\]

In our setting, constrained systems are those shift spaces which can be realized by walks on some labeled graph.

\begin{definition}
A shift space  $X\subseteq \Sigma^\Z$ is called a \textit{constrained system} (or a sofic shift) if there exists a finite directed graph $G=(V,E)$ and a labeling function $L:E\to \Sigma$ such that 
\[X=X_G\eqdef\set*{\parenv*{L(e_i)}_{i\in \Z}; (e_i)_{i\in \Z}  \text{ is a bi-infinite directed path in }G }.\] 
\end{definition}

A labeled graph $G=(V,E,L)$ is called irreducible if any two vertices are connected by  a directed path. An irreducible graph is called primitive if the greatest common divisor of all cycle lengths is $1$. It is well known (e.g., see~\cite[Theorem 4.5.8]{lind2021introduction}) that an irreducible graph is primitive if and only if there exists $n\in \N$ such that for any two vertices $v,v'\in V$ there exists a directed path of length $n$ from $v$ to $v'$. 
\begin{definition} A constrained system $X\subseteq \Sigma^\Z $ is called irreducible (respectively:  primitive), if there exists an irreducible (respectively: primitive) labeled graph $G$ such that $X=X_G$.
\end{definition}

A special family of constrained systems of particular interest is the family of systems defined by a finite set of local constraints. These are referred to as \emph{systems of finite type} and are formally defined as follows:  

\begin{definition}
A constrained system  $X\subseteq \Sigma^\Z$ is said to be a system of finite type (SFT) if there exists some $m\in \N$ and a finite set  of forbidden words $\cF\subseteq \Sigma^m$  such that $X$ is the set of all bi-infinite sequences not containing any forbidden pattern from $\cF$. That is 
\[ X=X_{\cF}\eqdef\set*{\bx\in \Sigma^\Z ; \forall n\in \Z, \bx_{n}^{n+m-1}\notin \cF}.\] 
\end{definition}

The coding-theoretic literature on the covering radius of error-correcting codes is quite extensive (e.g., see~\cite{Cohen} and the many references within). However, as will become apparent, our setting is quite different since we are considering  sequences of codes of finite block lengths, associated with constrained systems. As mentioned above, these constrained systems are mathematically formulated and studied using bi-infinite sequences. Thus, we shall borrow tools from ergodic theory to apply to the problem at hand.

\section{The Covering Radius of a Constrained System}\label{sec:CovRad}

We begin with a definition of the covering radius of a set $B\subseteq \Sigma ^n$ with respect to another set $A\subseteq \Sigma^n$ under the Hamming metric.

\begin{definition}
Let $A,C\subseteq \Sigma ^n$, then the covering radius of $C$ with respect to  $A$ is defined to be
\[  R(C,A)\eqdef \min\set*{r\in \N\cup\set{0} ; A\subseteq \bigcup_{\ox\in C}B_r(\ox) }=\max_{\oy\in A}\min_{\ox\in C}d(\ox,\oy).\]
If $A=\Sigma^n$ then $R(C,A)$ is just the regular covering radius of $C$, and is denoted by $R(C)$.
\end{definition}

For constrained systems $X,Y\subseteq \Sigma^\Z$ we define the asymptotic covering radius of $X$ with respect to $Y$ to be the asymptotic normalized covering  radius of $n$-tuples from $X$ with respect to  $n$-tuples from $Y$.
\begin{definition} Let $X,Y\subseteq\Sigma^\Z$ be shift spaces, then we define
\begin{equation} \label{eq:RadDef}
    R(X,Y)\eqdef \liminf_{n\to \infty}\frac{R(\sB_n(X),\sB_n(Y))}{n},
\end{equation}
where we remind that $\sB_n(X)$ and $\sB_n(Y)$ are the subwords of length $n$ from $X$ and $Y$ respectively.
\end{definition}

In a typical coding-theoretic framework, the covering radius is considered as a property of a single code in the Hamming space of a finite length $n$. A constrained system on the other hand may be associated with a sequence of codes, which are the sets of constrained words of fixed lengths. The covering radius of the constrained system, as defined above, is in fact the asymptotic value of the (normalized) covering radii of this corresponding sequence of codes.

We remark that throughout the most of this work we take $Y$ to be the whole space $\Sigma^\Z$, however the results stated hold for general constrained systems. An immediate question that comes up when considering our definition of the covering radius, is whether the limit from~\eqref{eq:RadDef} exists. We shall now show that the answer is yes when $X$ or $Y$ are primitive. For our proof, we recall the following generalization of Fekete's Lemma, for nearly subadditive sequences.

\begin{lemma}[De Bruijn–Erdős Condition \cite{de1951some,furedi2020nearly}] 
\label{lem:Fekete}
Let $(a_n)_{n\in\N}$ be a sequence of non-negative real numbers  satisfying for all $m,n\in \N$ 
\[a_{n+m}\leq a_n+a_m+f_{n+m},\]
where $(f_k)_{k\in\N}$ is a sequence of numbers such that 
$\sum_{k=1}^\infty\frac{f_k}{k^2}<\infty$.
Then the sequence $(\frac{a_n}{n})_{n\in\N}$ converges.
\end{lemma}


\begin{prop} \label{prop:LimExists}
Assume that $X, Y\subseteq \Sigma^\Z$ are constrained systems.  If $X$ or $Y$ are primitive, then the $\liminf$ in the definition of $R(X,Y)$ is actually a limit:
\[R(X,Y)=\lim_{n\to\infty}\frac{R(\sB_n(X),\sB_n(Y))}{n}.\]
\end{prop}

{\color{red} }
\begin{IEEEproof}
We begin with the case where $X$ is primitive, and is presented by the primitive finite labeled graph $G=(V,E,L)$. Since $G$ is primitive, there exists a sufficiently large $N$ such that any two vertices are connected by a directed path of length $N$. This implies that for any two words $\ou,\ov\in \sB(X)$ there exists $\ow\in \Sigma^N$ such that $\ou\,\ow\,\ov\in \sB(X)$.
For $n\in \N$ let us denote $a_n=R(\sB_n(X),\sB_n(Y))$. We show that for all $n>N$ and $m\geq 1$, we have $a_{n+m}\leq a_n+a_m+N$. In order to show the desired inequality it is sufficient that we prove the following statement: given $\oy\in \sB_{n+m}(Y)$ there exists $\ox \in \sB_{n+m}(X)$ such that $d(\ox,\oy)\leq a_n+a_m+N$. From the definitions of $a_m$ and $a_{n-N}$ there exists $\ou\in \sB_m(X)$ and $\ov\in \sB_{n-N}(X)$ such that 
\[d(\ou,\oy_0^{m-1})\leq a_m \quad \text{and} \quad  d(\ov,\oy_{m+N}^{m+n-1})\leq a_{n-N}.\]

We also observe immediately from the definition that $(a_k)_{k\in\N}$ is a non-decreasing sequence and therefore $a_{n-N}\leq a_n$. Let $\ow\in \Sigma^N$ be such that $\ox=\ou\,\ow\,\ov\in\sB_{n+m}(X)$. From the properties of the Hamming metric
\begin{align*}
    d(\ox,\oy)&= d(\ov,\oy_{0}^{m-1})+ d(\ow,\oy_{m}^{m+N-1})+d(\ou,\oy_{m+N}^{m+n-1})\\
    &\leq a_m+N+a_{n-N}\leq a_m+a_n+N.
\end{align*}
When $n\geq 1$ and $m>N$ a symmetric analysis follows. Since the remaining cases, i.e., both $m,n\leq N$, comprise of a finite number of cases, let $c=\max\parenv*{N,\max\set*{a_{m+n}-a_n-a_m ;n,m\leq N}}$. Let $(f_n)_{n\in\N}$ be defined by $f_n=c$ for all $n$. For all $m,n\in \N$ we have $a_{n+m}\leq a_n+a_m+f_{n+m}$, and therefore by Lemma~\ref{lem:Fekete}, the sequence $(\frac{a_n}{n})_{n\in\N}$ converges, as desired.

We now turn to the the second case of the proposition, where $Y$ is primitive. As in the first part, by the primitivity of $Y$, there exists an $N$ such that for any word $\ou\in \sB(Y)$ there exists $\ou'\in \Sigma^N$ such that $\ou\,\ou'\ou\in \sB(Y)$. Now let $\ou\in \sB_m(Y)$ be a deep hole, namely, a word such that $R(\sB_m(X),\sB_m(Y))=\min\set*{d(\ox,\ou); \ox\in \sB_m(X)}$. Denote $m'=m+N$. For $n\geq m'$, we consider the word $\oy\in \sB_n(Y)$ defined by 
\[\oy=(\overset{\floor*{\frac{n}{m'}}\text{ times}}{\overbrace{\ou,\ou',\ou,\ou',\dots,\ou,\ou'},}\oy'),\]
where $\oy'$ is an arbitrary suffix such that $\oy\in \sB_n(Y)$. We remark that it is possible to find $\oy'$, for example by taking a prefix of $u$. For any $\ox\in \sB_n(X)$ we have 
\begin{align*}
    d(\ox,\oy)&\geq \sum_{i=0}^{\floor*{\frac{n}{m'}}-1}d\parenv*{\ox_{i m'}^{(i+1)m'-1},\oy_{i m'}^{(i+1)m'-1}}=\sum_{i=0}^{\floor*{\frac{n}{m'}}-1}d\parenv*{\ox_{i m'}^{(i+1)m'-1},(\ou,\ou')}\\
    &\geq \sum_{i=0}^{\floor*{\frac{n}{m'}}-1}d\parenv*{\ox_{i m'}^{(i+1)m'-1-N},\ou} \geq \floor*{\frac{n}{m'}}R(\sB_m(X),\sB_m(Y))=\floor*{\frac{n}{m+N}}R(\sB_m(X),\sB_m(Y)).
\end{align*}
This proves that $R(\sB_n(X),\sB_n(Y)) \geq \floor{\frac{n}{m+N}}R(\sB_m(X),\sB_m(Y))$. Taking the limit we have 
\begin{align*}
    \liminf_{n\to\infty} \frac{R(\sB_n(X),\sB_n(Y))}{n}\geq \liminf_{n\to\infty} \frac{\floor*{\frac{n}{m+N}}R(\sB_m(X),\sB_m(Y))}{n}=\frac{R(\sB_m(X),\sB_m(Y))}{m+N}. 
\end{align*}
This holds for all $m\in \N$, which proves that 
\[\liminf_{n\to\infty} \frac{R(\sB_n(X),\sB_n(Y))}{n}\geq \limsup_{m\to\infty} \frac{R(\sB_m(X),\sB_m(Y))}{m+N}=\limsup_{m\to\infty} \frac{R(\sB_m(X),\sB_m(Y))}{m},\]
and the limit exists.
\end{IEEEproof}

\begin{remark}\label{rem:LimIsInf}
From the proof of Proposition~\ref{prop:LimExists} it follows that if $Y=\Sigma^\Z$ then
\[R(X,Y)=\lim_{n\to\infty} \frac{R(\sB_n(X),\Sigma^n)}{n}=\sup_{n\in \N}\frac{R(\sB_n(X),\Sigma^n)}{n}.\]
\end{remark}

We shall use the following as a running example throughout the paper. 

\begin{example} \label{ex:0kRLL}
Consider the binary alphabet, $[2]\eqdef\set{0,1}$. Let $X_{0,k}\subseteq [2]^\Z$ be the $(0,k)$-RLL system, which comprises of all the binary sequences that do not contain $k+1$ consecutive zeros. That is, $X_{0,k}=X_{\cF}$, where $\cF=\set*{\ozero_{k+1}=(0,\dots,0)}\subseteq [2]^{k+1}$. We claim that
\[R(X_{0,k},[2]^\Z)=\frac{1}{k+1}.\]
Indeed, by Remark~\ref{rem:LimIsInf} 
\begin{align*}
    R(X_{0,k},[2]^\Z)&\geq \frac{R(\sB_{k+1}(X_{0,k}),[2]^{k+1})}{k+1}=\frac{R\parenv{[2]^{k+1}\setminus\set{\ozero_{k+1}},[2]^{k+1}}}{k+1}=\frac{1}{k+1}.
\end{align*}
We now show that the obtained lower bound is tight. Let $\oy\in[2]^n$ be any binary word. Consider $\ox$ given by 
\[\ox_i\eqdef \begin{cases} \oy_i & i\bmod (k+1) \neq 0,\\
1 & i\bmod (k+1) = 0.
\end{cases}\]
Clearly $\ox$ does not contain any subword of $k+1$ consecutive zeros and therefore $\ox\in \sB_n(X_{0,k})$. Since $d(\ox,\oy)\leq \ceil{\frac{n}{k+1}}$ we conclude that $\frac{1}{n} R(\sB_n(X_{0,k}),[2]^n)\leq \frac{1}{n}\ceil{\frac{n}{k+1}}$, and by taking the limit,
\[ R(X_{0,k},[2]^\Z)=\lim_{n\to\infty}\frac{R(\sB_n(X_{0,k}),[2]^n)}{n} \leq \frac{1}{k+1}.\]
\end{example}

Using a ball-covering argument, we provide a simple lower-bound on the covering radius  in term of the capacities of the systems. We recall that $H_q:[0,1]\to[0,1]$ denotes the $q$-ary entropy function defined by
\[ H_q(x) \eqdef x\log_q (q-1) - x\log_q (x) - (1-x)\log_q (1-x),\]
and for continuity, $H_q(0)\eqdef 0$ as well as $H_q(1)\eqdef\log_q(q-1)$. We also use $H_q^{-1}:[0,1]\to[0,1-\frac{1}{q}]$ to denote its inverse function.

\begin{prop}\label{prop:LB-Union}
Let $X,Y\subseteq \Sigma^\Z$ be constrained systems with capacities $h(X)\leq h(Y)$, respectively, and let us denote $\abs{\Sigma}=q$. Then
\[
    R(X,Y)\geq H_q^{-1}(h(Y)-h(X)).
\]
\end{prop}
\begin{IEEEproof}
 If $R(X,Y)\geq 1-\frac{1}{q}$ the claim follows immediately from the definition of $H_q^{-1}$. Thus, we assume that $R(X,Y)< 1-\frac{1}{q}$. Let $V_{r,n,q}$ denote the size of a ball of radius $r$ in $\Sigma^n$ with respect to the Hamming metric (which is invariant to the choice of the center), and let us denote $\rho_n\eqdef \frac{1}{n}R(\sB_n(X),\sB_n(Y))$. By the union bound, for any $n\in \N$, 
 \begin{equation}
     \label{eq:BallCovBound}
     |\sB_n(Y)|=\abs*{\sB_n(Y)\cap \parenv*{\bigcup_{\ox\in \sB_n(X)}B_{n \rho_n}(\ox)}}\leq \abs*{\sB_n(X)}\cdot V_{\rho_n n ,n,q}.
 \end{equation}
 By a standard use of Stirling's approximation (e.g., see~\cite[Chapter 3]{GurRudSud22}) it is well-known that
\begin{align*}
    V_{\rho n,n,q}=\sum_{i=0}^{\floor{\rho n}} \binom{n}{i}(q-1)^i\leq \begin{cases}
    q^{n H_q(\rho)} & \rho\in[0,1-\frac{1}{q}), \\
    q^n & \rho\in[1-\frac{1}{q},1].
    \end{cases}
\end{align*}
Since the limit defining the capacity $h(X)$ exists, by \eqref{eq:BallCovBound} and the continuity of $H_q$ we obtain
 \begin{align*}
     h(X)&=\lim_{n\to\infty}\frac{\log_q\abs*{\sB_n(X)}}{n}=\liminf_{n\to\infty}\frac{\log_q\abs*{\sB_n(X)}}{n}\\
     &\geq \liminf_{n\to\infty} \parenv*{\frac{\log_q|\sB_n(Y)|}{n}-H_q(\rho_n)}=h(Y)-H_q(R(X,Y)).
 \end{align*}
    By rearranging and employing $H_q^{-1}$ we conclude.
\end{IEEEproof}


\begin{example}
\label{ex:repShift} 
Fix $\Sigma=[q]\eqdef\set{0,\dots,q-1}$ and consider the repetition shift
\[X_{\rep}\eqdef \set*{(\dots,a,a,a,\dots); a\in [q]}.\]
Clearly, $X_{\rep}$ is the SFT defined by the set of forbidden patterns 
\[\cF=\set*{ab;a,b\in [q], a\neq b}\subseteq [q]^2.\] 
Since $h(X_{\rep})=0$, by Proposition~\ref{prop:LB-Union}
\[R(X_{\rep},[q]^\Z)\geq H_q^{-1}(1-0)=1-\frac{1}{q}.\]
On the other hand, for any $n\in\N$ and for any $\oy\in [q]^n$, it is clear that there exists at least one symbol $a\in [q]$ which appears in at least $\ceil{\frac{n}{q}}$ coordinates of $\oy$, and in particular $d(\oy,(a,\dots,a))\leq \floor{\frac{q-1}{q}n}$. This proves that
\[R(\sB_n(X_{\rep}),[q]^n)\leq\floor*{\frac{q-1}{q}n}.\]
Taking the limit and combining with the lower bound
\[R(X_{\rep},[q]^\Z)=1-\frac{1}{q}.\]
\end{example}

\begin{remark}
Example~\ref{ex:repShift} shows that the covering radius of a union of two constrained systems can be strictly smaller then the minimum of the covering radii. We note that $X_{\rep}=\bigcup_{a\in [q]}\set{\bx_a}$, where $\bx_a$ is the constant bi-infinite sequence of the symbol $a$. We note that $\set{\bx_a}$ is a constrained system and that $R(\set{\bx_a},[q]^\Z)=1$. Thus 
\[\min_{a\in [q]}R(\set{\bx_a},[q]^\Z)=1>1-\frac{1}{q}=R(X_\rep,[q]^\Z).\]
That is in contrast to  the capacity of constrained systems, where the capacity of a finite union is equal to the maximum of the capacities.
\end{remark}
At this point we have reached a curious situation. For the sake of illustrating it, fix the binary alphabet $\Sigma=[2]$. If we consider $X_{0,1}$, the $(0,1)$-RLL system from Example~\ref{ex:0kRLL}, then its capacity is known to be $h(X_{0,1})=\log_2((1+\sqrt{5})/2)\approx 0.694$, and we have shown that its covering radius (with respect to $[2]^\Z$) is $R(X_{0,1},[2]^\Z)=\frac{1}{2}$. However, in Example~\ref{ex:repShift} we have seen that the binary repetition shift, $X_{\rep}$, has the same covering radius $R(X_{\rep},[2]^\Z)=\frac{1}{2}$, but zero capacity, $h(X_{\rep})=0$. From a coding perspective, even though $\sB_n(X_{0,1})$ has exponentially more words than $\sB_n(X_{\rep})$, the worst-case covering scenario, namely, a deep hole, is asymptotically within the same distance from the constrained code.

Apart from the mathematical curiosity, having $R(X_{0,k},[2]^\Z)=R(X_{\rep},[2]^\Z)=\frac{1}{2}$ hinders (in these two example cases) the possibility of correcting channel errors in the QCC scheme described in Section~\ref{sec:intro}. This is because the error-correcting code needs to correct more erroneous positions than $\frac{1}{2}$ of the code length, which is impossible to do with a non-vanishing rate. We are therefore motivated to seek a different version of the covering radius of a constrained system, which takes into account the rarity of deep holes. 

As a final comment for this section, we would like to comment on the relation of $R(X,\Sigma^\Z)$ to the QCC framework. Since we are interested in asymptotics, assume that the sequence of error-correcting codes in the QCC scheme is $(C_n)_{n\in\N}$, where $C_n$ is of length $n$.  The expression $R(X,\Sigma^\Z)=\lim_{n\to\infty}\frac{1}{n}R(\sB_n(X),\Sigma^n)$ is an upper bound on the worse-case quantization error rate using a sequence of codes $(C_n)_{n\in\N}$, which is actually $\lim_{n\to\infty}\frac{1}{n}R(\sB_n(X),C_n)$.  The bound $R(X,\Sigma^\Z)$ is pessimistic twice: once for allowing deep holes to determine the covering radius, and twice, for assuming they reside in $C_n$. Since $\lim_{n\to\infty}\frac{1}{n}R(\sB_n(X),C_n)$ may be hard to compute and depends on the sequence of error-correcting codes, we may use $R(X,\Sigma^\Z)$ as an upper bound on the worse-case quantization error, which is independent of the sequence of codes.


\section{The Essential Covering Radius}\label{sec:AverageCov}

The covering radius that was studied in the previous section may be perhaps too pessimistic in the sense that it is determined by the worst-case quantization distance. In this section we study a different definition of the covering radius, which we call the essential covering radius. Given $\varepsilon>0$, the $\epsilon$-covering radius of a constraint system is, loosely speaking, the smallest $r$ such that $(1-\varepsilon)$-fraction the words in the space can be quantized to the constraint system. In what follows, we further generalize this to a probabilistic definition of the covering radius.

We begin by stating some basic definitions and well known results from ergodic theory.  For any finite alphabet $\Sigma$, we consider $\Sigma^\Z$ as a measurable space, together with the Borel $\Sigma$-algebra induced by the product topology on $\Sigma^\Z$. Similarly, any subshift $Y\subseteq \Sigma^\Z$ is considered as a measurable space.

\begin{definition}[Invariant and ergodic measures] 
Let $Y\subseteq \Sigma^\Z$ be a subshift. A probability measure $\mu$ on $Y$ is called \emph{shift invariant} if $\mu(T^{-1}B)=\mu(B)$ for any measurable set $B$. A shift-invariant measure $\mu$ is further said to be \emph{ergodic} if $T^{-1}B=B$ implies $\mu(B)=0$ or $\mu(Y\setminus B)=0$.
The set of shift-invariant probability measures on $Y$ is denoted by $M(Y)$, and the set of ergodic measures in $M(Y)$ is denoted by $M_{\cE}(Y)$. 
\end{definition}

For a measure $\mu\in M(Y)$ we denote by $\mu_n$ the marginal measure  of $\mu$  on the first $n$ coordinates, which is a probability measure on $\Sigma^n$. To avoid cumbersome notation, throughout this work we shall use $\Pro_\mu[A]$ in order to denote the measure $\mu(A)$, and $\bY$ for a random bi-infinite sequence on $Y$.  Throughout this article we use bold upper-case letters for bi-infinite sequences of random variables, not to be confused with non-bold capital letters used to denote constrained systems.  An important result that we use in our analysis is the following well-known ergodic 
theorem, which is a classical result in Ergodic Theory. The $L_2$ convergence in the ergodic theorem is due to von-Neumann and the almost-surely result is due to Birkhoff, both from 1931. A proof of this well-known classical theorem can be found in most standard introductory textbooks on ergodic theory, for instance~\cite[Chapter 3]{rudolph1990fundamentals}.


\begin{theorem}[{{The Ergodic Theorem, \cite{1931-Birkhoff, 1932-Neumann}}}]
\label{th:Ergodic Theorem}
Let $Y\subseteq \Sigma^\Z$ be a shift space, $\mu \in M_{\cE}(Y)$ be an ergodic measure, and $f\in L_1(\mu)$ be an integrable function. Then the sequence $(A_n)_{n\in \N}$ defined by
\[A_n=\frac{1}{n} \sum_{i=0}^{n-1}f\circ T^{-i}\]
converges almost-surely, and in $L_2$, to  $\intop f \cdot d \mu$. In particular, for all $\delta>0$
\[ \lim_{n\to \infty} \Pro_{\mu}\sparenv*{\abs*{A_n-\intop f\cdot d\mu}>\delta}=0.\]
\end{theorem}

We are now ready to define the essential covering radius.
    
\begin{definition}
For any real $\varepsilon>0$, two sets $A,C\subseteq \Sigma^n$, and $\eta$, a probability measure on $A$, we define 
\[R_\varepsilon(C,A,\eta)\eqdef\min\set*{r\in \N\cup\set{0} ; \eta \parenv*{A\cap \parenv*{\bigcup_{\ox\in C}B_r(\ox)}}\geq 1-\varepsilon}.\]
\end{definition}
    
We remark that when $\eta$ is the uniform measure on $A$, $R_\varepsilon(C,A,\eta)$ is the $\varepsilon$-covering radius of $A$, namely the smallest $r$ such that at least $(1-\epsilon)$-fraction of the words in $C$ are at distance at most $r$ from $A$, as desired.
    
\begin{definition}
Let $X,Y\subseteq \Sigma^\Z$ be constrained systems, and $\mu\in M_{\cE}(Y)$ be an ergodic measure.  We define the \emph{$\varepsilon$-covering radius} of $X$ with respect to $(Y,\mu)$  by
\[R_\varepsilon(X,Y,\mu)\eqdef  \liminf_{n\to \infty}\frac{R(\sB_n(X),\sB_n(Y),\mu_n)}{n},\]
and the \emph{essential covering radius} of $X$ with respect to $(Y,\mu)$ by
\[R_0(X,Y,\mu)\eqdef \lim_{\varepsilon\to 0}R_\varepsilon(X,Y,\mu).\]
\end{definition}

We comment that the limit in the previous definition exists due to the monotonicity of $R_\varepsilon(X,Y,\mu)$ in $\varepsilon$. We also observe that, trivially, the essential covering radius is upper bounded by the (worst-case) covering radius, and we have
\begin{equation}
    \label{eq:epsAvReg}
    R_{\varepsilon}(X,Y,\mu)\leq R_0(X,Y,\mu) \leq R(X,Y).
\end{equation}

We now review the examples of the repetition system (Example~\ref{ex:repShift}) and the $(0,k)$-RLL system (Example~\ref{ex:0kRLL}), considered in the previous section. 
    
\begin{prop}\label{prop:repAv}
Consider the $q$-ary repetition system $X_{\rep} \subseteq [q]^{\Z}$ from Example~\ref{ex:repShift}, and assume $Y=[q]^\Z$ is equipped with the uniform Bernoulli i.i.d measure, denoted by $\muni$.  Then the essential covering radius is equal to the covering radius, i.e.,
\[R_0(X_{\rep},[q]^\Z,\muni)=R(X_{\rep},[q]^\Z)=1-\frac{1}{q}.\]
\end{prop}
\begin{IEEEproof}
For a fixed $n\in \N$ and $i\in [q]$, let $Y_{n}^{(i)}$ be the random variable that counts the number of coordinates with the symbol $i$ in a random (uniformly distributed) word in $[q]^n$. By the law of large numbers, for any $\delta>0$ 
 \begin{equation}
     \label{eq:LLN1}
     \lim_{n\to \infty}\Pro_{\muni_n}\sparenv*{\bigcap_{i=0}^{q-1}\set*{\abs*{\frac{Y_n^{(i)}}{n}-\frac{1}{q}}<\delta}}=1.
\end{equation}
We also note that for a word $\oy\in [q]^n$ with each symbol $i\in[q]$ appearing in at least $(\frac{1}{q}-\delta)n$ coordinates, we have
\begin{equation}\label{eq:deltaDist}
     \min_{\ox\in \sB_n(X_{\rep})}d(\ox,\oy)\geq(q-1)\parenv*{\frac{1}{q}-\delta}n.
\end{equation}
For any $\varepsilon>0$, from \eqref{eq:LLN1} it follows that for sufficiently large $n$, any set in $[q]^n$, of probability at least $1-\varepsilon$, contains a word such that any $i\in[q]$ appears in at least $(\frac{1}{q}-\delta)n$ coordinates. Combining this with \eqref{eq:deltaDist} we conclude that for sufficiently large $n$
\[\frac{R_{\varepsilon}(\sB_n(X_{\rep}),[q]^n,\muni_n)}{n}\geq (q-1)\parenv*{\frac{1}{q}-\delta}.\]
It then follows that $R_{\varepsilon}(X_{\rep},[q]^\Z,\muni)\geq (q-1)(\frac{1}{q}-\delta) $, which is true for all $\delta>0$, and therefore 
\[R_0(X_{\rep},[q]^\Z,\muni)=\lim_{\varepsilon\to 0} R_{\varepsilon}(X_{\rep},[q]^\Z,\muni)\geq 1-\frac{1}{q}.\]
The upper bound follows trivially from Example~\ref{ex:repShift}, and we have 
\[R_0(X_{\rep},[q]^\Z,\muni)=R(X_{\rep},[q]^\Z)=1-\frac{1}{q}.\]
\end{IEEEproof}

\begin{remark}
Since \eqref{eq:epsAvReg} is true for all ergodic measures, we can lower bound the covering radius as follows, 
\begin{equation}
    \label{eq:AvCombBound}
    R(X,Y)\geq \sup_{
\mu\in M_\cE(Y)}R_0(X,Y,\mu).
\end{equation}
Proposition~\ref{prop:repAv} shows that in the case where $X=X_{\rep}$ is the repetition system and $Y=[q]^\Z$, \eqref{eq:AvCombBound} is tight, as equality holds for the uniform Bernoulli measure on $Y$.

We claim that~\eqref{eq:AvCombBound} is tight also in the case where $X=X_{0,k}$, the binary $(0,k)$-RLL system described in Example~\ref{ex:0kRLL}. To see this, consider $\mu=\delta_{\bzero}$, the Dirac measure of the all-zero sequence $\bzero \in Y=[2]^\Z$. Since $\bzero$ is a periodic point with period $1$ (with respect to the shift transformation), $\delta_{\bzero}$ is indeed an invariant measure, which is also ergodic, as it assigns a measure of $0$ or $1$ to any set. A simple calculation shows that for any $0<\varepsilon < 1$ and $n\in \N$, 
\[R_\varepsilon(\sB_n(X_{0,k}),[2]^n,\mu_n)=\min_{\ox \in \sB_n(X_{0,k})}d(\ox,\ozero_n)=\floor*{\frac{n}{k+1}},\]
which implies
\[R_0(X_{0,k},[2]^\Z,\mu)=\lim_{\varepsilon\to 0}R_\varepsilon(X_{0,k},[2]^\Z,\mu)=\lim_{\varepsilon\to 0}\liminf_{n\to\infty} \frac{R_\varepsilon(\sB_n(X_{0,k}),[2]^n,\mu_n)}{n}=\frac{1}{k+1}=R(X,Y).\]
We conjecture that the bound \eqref{eq:AvCombBound} is tight in general, and leave it as a direction for future work.
\end{remark}

As we have seen, the repetition system, whose capacity is zero, has the same covering radius and essential covering radius. The $(0,k)$-RLL system has positive capacity. While its covering radius is $\frac{1}{k+1}$, the following theorem shows its essential covering radius decays exponentially fast with $k$, in stark contrast to the repetition system.

\begin{theorem}\label{th:0k} 
Let $X_{0,k}\subseteq [2]^\Z$ be the  $(0,k)$-RLL system from Example~\ref{ex:0kRLL}, and let   $Y=[2]^\Z$ be equipped with the uniform Bernoulli i.i.d measure $\muni$. Then 
\[ R_0(X_{0,k},[2]^\Z,\muni)=\frac{1}{2(2^{k+1}-1)}. \]
\end{theorem}

\begin{IEEEproof}
For a word $\oy\in [2]^n$ and an integer $i\geq k+1$, let us denote by $S_i(\oy)$ the number of appearances of the pattern $1\ozero_i 1\in [2]^{i+2}$ in $\oy$. We also denote by $M(\oy)$ the number of coordinates $j\in [n]$ that are not part of a pattern of the form $1\ozero_i 1$ in $\oy_i$, for any $i\geq k+1$. The key observation is the following inequality, asserting that for all $\ell \geq k+1$ we have
\begin{equation}\label{eq:0kdistY}
    \sum_{i=k+1}^\ell  \floor*{\frac{i}{k+1}}S_i(\oy)\leq \min_{\ox\in \sB_n(X_{0,k})} d(\ox,\oy)\leq  M(\oy)+ \sum_{i=k+1}^\infty  \floor*{\frac{i}{k+1}}S_i(\oy).
\end{equation}
Indeed, we note that for any  $\ox\in\sB_n(X_{0,k})$ and for any instance of the pattern  $\oy_{m}^{m+i+1}=1\ozero_i1$, we have that $\ox_{m+1}^{m+i}$ and $\oy_{m+1}^{m+i}$ must differ in at least $\floor{\frac{i}{k+1}}$ places 
since $\ox$ does not contain $k+1$ consecutive zeros. We further observe that patterns of the form $1\ozero_i 1$ in $\oy$ do not overlap in zeros. Thus we obtain,
\[
    d(\ox,\oy)\geq \sum_{i=k+1}^{\infty} \floor*{\frac{i}{k+1}} S_i(\oy) \geq \sum_{i=k+1}^{\ell} \floor*{\frac{i}{k+1}} S_i(\oy),
\]
and the lower-bound follows by taking the minimum over all $\ox\in \sB_n(X_{0,k})$. On the other hand, in order to prove the upper bound, it suffices to construct a word $\ox\in \sB_n(X_{0,k})$ satisfying
\begin{equation}\label{eq:closeX}
    d(\ox,\oy)\leq M(\oy)+ \sum_{i=k+1}^{\infty} \floor*{\frac{i}{k+1}} S_i(\oy).
\end{equation}
We construct $\ox$ as follows: for any $i\geq k+1$ and any instance of the pattern $1\ozero_i 1$ in $\oy_m^{m+i+1}$, we set 
\[\ox_m^{m+i+1}=1  \overset{\floor{\frac{i}{k+1}}(k+1)}{\overbrace{\ozero_k1 \ozero_k 1\cdots \ozero_k 1}} \overset{i \bmod (k+1)}{\overbrace{0\cdots 0}} 1.\]
In the remaining coordinates we define $\ox$ to be the same as $\oy$ except the coordinates counted by $M(\oy)$ which we set to $1$. From the construction of $\ox$, the longest run of zeros it contains is at most $k$, which implies $\ox\in \sB_n(X_{0,k})$. Thus, $\ox$ satisfies~\eqref{eq:closeX} as desired. 

For any $\ell \geq k+1$ (including $\ell =\infty$) we define $f_\ell : [2]^\Z\to \R$ to be
\[f_\ell (\by)=\begin{cases}
\floor*{\frac{i}{k+1}} & \by_{0}^{i+1}=1\ozero_i 1 \text{ for }k+1\leq i \leq \ell ,\\
0 & \text{otherwise,}
\end{cases}\]
and we denote 
\[E_\ell\eqdef\intop f_\ell \cdot d\muni.\]
A simple calculation shows that for any finite $\ell\in \N$ 
\[E_\ell=\sum_{i=k+1}^{\ell} \floor*{\frac{i}{k+1}}\frac{1}{2^{i+2}},\]
and by the monotone convergence theorem $E_\ell\to E_\infty$ as $\ell\to\infty$, since $f_\ell$ (monotonically) converges to $f_\infty$ pointwise.

Let $\oY_n$ be a random word of length $n$ distributed according to $\muni_n $, and let $\bY$ be a random sequence distributed according to $\muni$. We note that for any $\ell\leq n-2$, \[\sum_{j=0}^{n-\ell-2}f_\ell\circ T^{j}(\bY)= \sum_{i=0}^\ell\floor*{\frac{i}{k+1}}S_i(\bY_0^{n-1}).\]
Combining with \eqref{eq:0kdistY}, for a fixed $\ell$ and $\delta>0$, and for all sufficiently large $n$,
\begin{align*}
    \Pro_{\muni_n}\sparenv*{\min_{\ox_n\in \sB_n(X_{0,k})}\frac{d(\ox_n,\oY_n)}{n}>E_{\ell}-\delta}&= \Pro_{\muni}\sparenv*{\min_{\ox_n\in \sB_n(X_{0,k})}\frac{d(\ox_n,\bY_{0}^{n-1})}{n}>E_{\ell}-\delta}\\
    &\geq \Pro_{\muni} \sparenv*{\frac{1}{n}\sum_{i=k+1}^\ell \floor*{\frac{i}{k+1}}S_i(\bY_0^{n-1})>E_\ell-\delta}\\
    &= \Pro_{\muni} \sparenv*{\frac{1}{n}\sum_{j=0}^{n-\ell-2} 
    f_\ell\circ T^{j}(\bY)>E_\ell-\delta}\xrightarrow[n\to\infty]{}1,
\end{align*}
where the convergence to $1$ is due to the ergodic theorem (see Theorem~\ref{th:Ergodic Theorem}). This proves that for any $\varepsilon\in (0,1)$, $\delta>0$, and $\ell\geq k+1$, for sufficiently large $n$
\[\frac{R_{\varepsilon}(\sB_n(X_{0,k}),\sB_n(Y),\muni)}{n}> E_\ell-\delta,\]
and therefore, 
\[R_\varepsilon(X_{0,k},[2]^\Z,\muni)\geq \lim_{\substack{\ell\to\infty\\
\delta \to 0}}E_\ell-\delta =E_\infty.\]

For the upper bound, we use a similar technique. We first note that the sequence of random variables $(\frac{1}{n}M(\bY_0^{n-1}))_{n\in\N}$ converges in probability to $0$. We note that for any $N\in\N$, and a word  $\oy$ of length $n$, if $M(\oy)> 2N$ then $\oy_0^{N-1}=\ozero$ or $\oy_{n-N}^{n-1}=\ozero$. Thus for any $\delta>0$
\begin{align*}
    \Pro_{\muni}\sparenv*{\abs*{\frac{1}{n}M(\bY_0^{n-1})-0}>2\delta}&\leq \Pro_{\muni}\sparenv*{\bY_0^{\floor{\delta n}}=\ozero}+\Pro_{\muni}\sparenv*{\bY_{n-\floor{\delta n}}^{n-1}=\ozero}=\frac{2}{2^{\floor{\delta n}}}\xrightarrow[n\to\infty]{}0.
\end{align*}

We similarly note that for any $n$, 
\[\sum_{j=0}^{n-1}f_\infty\circ T^{j}(\bY)\geq \sum_{i=0}^\infty \floor*{\frac{i}{k+1}}S_i(\bY_0^{n-1}).\] 
Combining with \eqref{eq:0kdistY}, for all $\delta>0$
\begin{align*}
    \Pro_{\muni_n}\sparenv*{\min_{\ox_n\in \sB_n(X_{0,k})}\frac{d(\ox_n,\oY_n)}{n}<E_{\infty}+\delta}&\geq \Pro_{\muni}\sparenv*{\frac{1}{n}\parenv*{M(\bY_0^{n-1})+\sum_{i=k+1}^\infty \floor*{\frac{i}{k+1}}S_i(\bY_0^{n-1})}<E_{\infty}+\delta}\\
    &\geq \Pro_{\muni}\sparenv*{\frac{1}{n}\parenv*{M(\bY_0^{n-1})+\sum_{j=0}^{n-1} 
    f_\infty\circ T^{j}(\bY)}<E_{\infty}+\delta}\xrightarrow[n\to\infty]{}1,
\end{align*}
where again, the convergence to $1$ follows from the ergodic theorem and from the convergence in probability of $\frac{1}{n}M(\bY_0^{n-1})$ to $0$. As before, this proves that $R_\varepsilon(X_{0,k},[2]^\Z,\muni)\geq E_\infty$ for all $\varepsilon>0$, and by the lower bound $R_\varepsilon(X_{0,k},[2]^\Z,\muni)= E_\infty$. In particular, 
\[R_0(X_{0,k},[2]^\Z,\muni)=\lim_{\varepsilon\to 0}R_\varepsilon(X_{0,k},[2]^\Z,\muni)=E_\infty.\]

In order to complete the proof it only remains to compute $E_\infty$: 
\begin{align*}
    E_\infty&=\sum_{i=k+1}^\infty \floor*{\frac{i}{k+1}}\frac{1}{2^{i+2}}= \sum_{j=1}^{\infty} j \sum_{i=j(k+1)}^{(j+1)(k+1)-1}\frac{1}{2^{i+2}}=\frac{2^{k+1}-1}{2^{k+2}}\sum_{j=1}^\infty \frac{j}{2^{j(k+1)}} \\
    &= \frac{2^{k+1}-1}{2^{k+2}}\cdot\frac{2^{k+1}}{(2^{k+1}-1)^2}=\frac{1}{2(2^{k+1}-1)}.
\end{align*}
\end{IEEEproof} 

 It is desirable to have alternative expressions for the essential covering radius, which could assist in calculating or estimating  its value. Inspired by tools used in the proof of Theorem~\ref{th:0k}, we give an equivalent ergodic-theoretic characterization. 

\begin{definition}
 Let $X,Y\subseteq \Sigma^\Z$ be shift spaces, we consider $X\times Y$ as shift space, with the left shift acting 
 as $T(\bx,\by)=(T\bx,T\by)$. For an ergodic measure $\mu\in M_{\cE}(Y)$, an extension of $\mu$ over $X\times Y$ is a shift-invariant measure $\nu$ on the product space $X\times Y$ whose $Y$-marginal is $\mu$. Namely, $\nu$ satisfies that for any measurable $A\subseteq Y$
\[\nu(X\times A)=\mu(A).\]
 An extension on $X\times Y$ is said to be ergodic if it is an ergodic measure with respect to the shift transformation on the product space. We let $M(X,Y,\mu)$ denote the set of all extensions of $\mu$, and $M_{\cE}(X,Y,\mu)$ denote the set of all ergodic extensions in $M(X,Y,\mu)$. 
\end{definition}

In the following proposition, we provide an upper bound on the essential covering radius, that holds with no further assumptions on $X$ and $Y$. 

\begin{prop}\label{prop:JoinBound}
Let $X, Y\subseteq \Sigma^\Z$ be shift spaces, and $\mu\in M_{\cE}(Y)$. Then 
\[
    R_0(X,Y,\mu)\leq \inf \set*{\Pro_\nu[\bX_0\neq \bY_0]; \nu\in M_{\cE}(X,Y,\mu)},
\]
where $\bX_0, \bY_0$ are the random variables defined by the projections of two random sequences $\bX$ and $\bY$ on the $0$ coordinate. 
\end{prop}

\begin{IEEEproof}
Let $\nu$ be an extension in $M_{\cE}(X,Y,\mu)$, and let $0<\varepsilon<1$ and $\delta>0$ be arbitrarily small numbers. It is sufficient to prove that 
\[R_\varepsilon(X,\Sigma^\Z,\mu)\leq \Pro_\nu[\bX_0\neq \bY_0]+\delta.\]
We consider the function $f:X\times Y\to \set{0,1}$, defined to be the indicator function of the event $\set*{\bX_0\neq \bY_0}$. Clearly 
\[\intop f \cdot d\nu=\Pro_{\nu}[\bX_0\neq \bY_0].\] 
By Theorem~\ref{th:Ergodic Theorem}, for sufficiently large $n$,
\begin{equation}
    \label{eq:JoinBound1}
    \Pro_{\nu}\sparenv*{\abs*{\frac{1}{n}\sum_{i=0}^{n-1}f \circ {T^n}- \intop f \cdot d \nu}>\delta}<\varepsilon.
\end{equation}
We also observe that 
\begin{equation}
    \label{eq:JoinBoind2}
    \sum_{i=0}^{n-1}f \circ {T^n}= d\parenv*{\bX_0^{n-1},\bY_0^{n-1}},
\end{equation}
where $d$ is the Hamming distance. Let us denote
\[s_\delta\eqdef\Pro_{\nu}[\bX_0\neq \bY_0]+\delta \quad \text{and}\quad B_{s_\delta}(\sB_n(X))\eqdef \sB_n(Y)\cap \parenv*{\bigcup_{\ox\in \sB_n(X)}B_{s_\delta}(\ox)}.\] 
Combining \eqref{eq:JoinBound1} and \eqref{eq:JoinBoind2} with the law of total probability we obtain
\begin{align*}
    1-\varepsilon&<\Pro_{\nu}\sparenv*{d\parenv*{\bX_0^{n-1},\bY_0^{n-1}} \leq n\cdot s_\delta}\\
    &=\sum_{\oy\in \sB_n(Y)}\Pro_{\nu}\sparenv*{d\parenv*{\bX_0^{n-1},\bY_0^{n-1}} \leq n\cdot s_\delta  ~\bigg|~\bY_{0}^{n-1}=\oy}\Pro_{\nu}\sparenv*{\bY_{0}^{n-1}=\oy}\\
    &=\sum_{\oy\in B_{s_\delta}(\sB_n(X))}\Pro_{\nu}\sparenv*{d\parenv*{\bX_0^{n-1},\bY_0^{n-1}} \leq n\cdot s_\delta  ~\bigg|~\bY_{0}^{n-1}=\oy}\Pro_{\nu}\sparenv*{\bY_{0}^{n-1}=\oy}\\
    &\leq \sum_{\oy\in B_{s_\delta}(\sB_n(X))}\Pro_{\nu}\sparenv*{\bY_{0}^{n-1}=\oy}=\sum_{\oy\in B_{s_\delta}(\sB_n(X))}\Pro_{\mu}\sparenv*{\bY_{0}^{n-1}=\oy}\\
    &=\Pro_{\mu_n}\sparenv*{\sB_n(Y)\cap \parenv*{\bigcup_{\ox\in \sB_n(X)}B_{s_\delta}(\ox)}}.
\end{align*}
This shows that 
\[R_\varepsilon(X,Y,\mu)\leq s_\delta=\Pro_{\nu}[\bX_0\neq \bY_0]+\delta, \]
and therefore completes the proof.
\end{IEEEproof}

The following proposition shows that the minimization problem from the upper bound of Proposition~\ref{prop:JoinBound} over the set of ergodic extensions $M_{\cE}(X,Y,\mu)$ is in fact equivalent to a minimization over the set of all invariant extensions $M(X,Y,\mu)$. In order to show that, we shall require the ergodic decomposition theorem. Let $Z$ be a compact metric space equipped with the Borel $\sigma$-algebra and $T:Z\to Z$ be the a continuous function. We consider the space of $T$-invariant measures on $Z$, $M (Z)$, as a measurable space with the $\Sigma$-algebra induced by the weak-$*$ topology.

\begin{lemma}[{{Ergodic Decomposition, \cite[Theorem 4.8]{dani2012ergodic}}}] Let $Z$ be a compact metric space, $T:X\to X$ be a continuous map, and $\mu\in M (Z)$ be an invariant measure. Then there exists a unique probability measure $P_\mu$ on $M (Z)$, supported on the set of ergodic measures $M_{\cE}(Z)$, such that for any measurable set $E\subseteq Z$, 
\[\mu(E)=\intop_{M_{\cE}(Z)}\nu(E) dP_{\mu}(\nu).\]
\end{lemma}

\begin{prop} \label{prop:InfAll}
Let $X,Y\subseteq \Sigma^\Z$,  and let $\mu\in M_{\cE}(Y)$ be a shift-invariant ergodic measure. Then 
\[\inf\set*{\Pro_{\nu}[\bX_0\neq \bY_0]; \nu \in M_{\cE}(X,Y,\mu)}=\inf\set*{\Pro_{\nu}[\bX_0\neq \bY_0]; \nu \in M(X, Y,\mu)}.\] 
\end{prop}

\begin{IEEEproof} 
Let us denote $m\eqdef \inf\set*{\Pro_{\nu}[\bX_0\neq \bY_0]; \nu \in M_{\cE}(X,Y,\mu)}$. The inequality 
\[m\geq \inf\set*{\Pro_{\nu}[\bX_0\neq \bY_0]; \nu \in M (X,Y,\mu)}\]
is trivial as $M_{\cE}(X,Y,\mu) \subseteq M(X,Y,\mu)$. For the other direction, we are required to show that for all $\eta \in  M(X,Y,\mu)$ 
\[\Pro_{\eta}[\bX_0\neq \bY_0]\geq m.\]

Let $P_{\eta}$ be the ergodic decomposition of $\eta$. Namely $\eta=\intop_{ M_{\cE}(X\times Y)}\nu \cdot dP_{\eta}(\nu)$, and the set of ergodic measures has full $dP_{\eta}$-measure. We start by showing that $P_{\eta}$ is supported on $M_{\cE}(X,Y,\mu)$. Assume to the contrary that
\[P_{\eta}\sparenv*{\set*{\nu\in  M_\cE(X\times Y) ; \nu_Y\neq \mu}}>0.\]
Since the Borel $\Sigma$-algebra on $Y$ is countably generated, there exists a measurable set $E\subseteq Y$ and $n\in \N$ such that at least one of the sets $A_{\pm}$ defined by
\[A_{\pm}\eqdef \set*{\nu \in  M_{\cE}(X\times Y) ;  \pm\nu_Y(E)\geq \pm\mu(E)+\frac{1}{n}},  \] 
has a positive $P_{\eta}$-measure. Without loss of generality, assume $P_{\eta}(A_+)>0$. We define a new measure $\eta'$ by   
\[\eta'(A)=\frac{1}{P_{\eta}(A_+)}\intop_{A_+}\nu(A)\cdot dP_{\eta}(\nu).\]
We observe that $\eta'$ is a shift-invariant measure (as an integral over shift-invariant measures). We also note that $\eta'_Y\ll \eta_Y$ (that is, $\eta'_Y$ is absolutely continuous with respect to $\eta_Y$) as any null set with respect to $\eta$ is obviously a null set with respect to $\eta'$. It is well known that any invariant probability measure which is  absolutely continuous with respect to an invariant ergodic probability measure must be equal to it (e.g see \cite[Remark 1, p.153]{walters2000introduction}). 
We recall that $\eta_Y=\mu$ is ergodic, and therefore $\eta_Y=\eta_Y'$. This is a contradiction as from the definition of $A_+$ and $\eta'$ we have
\begin{align*}
    \mu(E)=\eta_Y(E)&=\eta'_Y(E)=\eta'(X\times E)=\frac{1}{P_{\eta}(A_+)}\intop_{A_+}\nu(X\times E)\cdot dP_{\eta}(\nu)\\
    &\geq \frac{1}{P_{\eta}(A_+)}\intop_{A_+}\parenv*{\mu(E)+\frac{1}{n}} dP_{\eta}(\nu)=\mu(E)+\frac{1}{n}.
\end{align*}
The claim now follows, since from the definition of $m$ we have
\begin{align*}
     \Pro_{\eta}[\bX_0\neq \bY_0]&=\intop_{M_{\cE}(X\times Y)}\Pro_{\nu}[\bX_0\neq \bY_0] dP_{\eta}(\nu)\\
     &=\intop_{M_\cE(X,Y,\mu)}\Pro_{\nu}[\bX_0\neq \bY_0] dP_{\eta}(\nu)\geq \intop_{M_\cE(X,Y,\mu)} m\cdot dP_{\eta}(\nu)=m.
 \end{align*}
\end{IEEEproof}

We are now ready to prove the main result of the section: the upper bound given in Proposition~\ref{prop:JoinBound} is in fact tight, and it provides an exact characterization of the essential covering radius by a minimization problem over invariant extensions.
\begin{theorem}
\label{th:ErgdicCaracter}
Let $X,Y\subseteq \Sigma^\Z$ be constrained systems, and let $\mu\in M_\cE(Y)$ be an ergodic measure. Then 
\begin{align*}
    R_0(X,Y,\mu)&=\inf\set*{\Pro_{\nu}[\bX_0\neq \bY_0]; \nu\in M_{\cE}(X,Y,\mu)}\\
    &=\inf\set*{\Pro_{\nu}[\bX_0\neq \bY_0]; \nu\in M(X,Y,\mu)}.
\end{align*}
\end{theorem}

\begin{IEEEproof}
By Proposition~\ref{prop:InfAll} and Proposition~\ref{prop:JoinBound}, it is sufficient to prove that for any $\delta>0$, there exists an invariant extension $\nu\in M(X,Y,\mu)$ with 
\begin{equation}
    \label{eq:measureCond}
    \Pro_\nu[\bX_0\neq \bY_0]\leq R_{0}(X,Y,\mu)+\delta.
\end{equation}
We shall  prove the existence of such an extension using the compactness of the simplex of  probability measures $X\times Y$ with respect to the weak-$*$ topology. We fix $\delta>0$ and find $\varepsilon<\delta/4$ in $(0,1)$ such that 
\begin{equation}\label{eq:CloseToZeroR}
    R_\varepsilon(X,Y,\mu)\leq R_0(X,Y,\mu)+\frac{\delta}{4}.
\end{equation}
For any fixed word $\oy\in \sB_n(Y)$ we fix an arbitrary $\ox(\oy)\in \sB_n(X)$ which is closest to $\oy$ among the words in $\sB_n(X)$.
From the definition of $R_\varepsilon(X,Y,\mu)$, there exists a sequence of distinct integers $(n_k)_{k\in\N}$ such that 
\[\lim_{k\to \infty}\frac{R_\varepsilon(\sB_{n_k}(X),\sB_{n_k}(Y),\mu_{n_k})}{n_k}=R_\varepsilon(X,Y,\mu).\]
From the definition of $R_\varepsilon(\sB_{n_k}(X),\sB_{n_k}(Y),\mu_{n_k})$ we have that for sufficiently large $k$
\begin{equation}\label{eq:CloseToEps}
    \Pro_{\mu}\sparenv*{\frac{1}{n_k}d\parenv*{\ox(\bY_0^{n_k-1}),\bY_0^{n_k-1}}<R_\varepsilon(X,Y,\mu)+\frac{\delta}{4}}>1-\varepsilon.
\end{equation}

For a fixed $k$, we define a map $f_k:Y\to X$ as follows: for any $\ox\in \sB_{n_k}(X)$ we fix some $\bx\in X$ such that $\bx_0^{n_k-1}=\ox$. We then define $f_k(\by)=\bx$, where $\bx$ is the sequence in $X$ corresponding to $\ox\parenv{\by_0^{n_k-1}}$.
Clearly $f_k$ is measurable since $f_k(\by)_m$ depends on finitely many coordinates of $\by$. We now consider the measure $\nu'_k$ on $X\times Y$, defined as the pushforward of $(f_k,\mathrm{Id}):Y\to X\times Y$. We also define  $\nu_k$ to be the measure obtained by averaging of $\nu_k'$ along the action of the shift, which is given by
\[\nu_=\frac{1}{n_k}\sum_{i=0}^{n_k-1}T^i_*\nu_k',\]
where $T^i_*\nu_k'$ is the pushforward of $\nu_k'$ with the $i$th-shift, defined by $T^i_*\nu_k'(A)=\nu_k'(T^{-i}A)$. We note that the $Y$-marginal of $\nu_k'$ is $\mu$ since $(f_k,\mathrm{Id})$ is the identity on the $Y$-coordinate. Since $\mu$ is shift invariant it follows that the  $Y$-marginal of  $\nu_k$  is also $\mu$.

By the compactness of the set of probability measures on $X\times Y$ with respect to the weak-$*$ topology, there exists a convergent subsequence $(\nu_{k_l})_{l\in\N}$, which by abuse of notation we denote by $(\nu_k)_k$. Let $\nu$ denote the weak-$*$ limit of $(\nu_k)_k$. Since the projection of a measure to its marginal is continuous with respect to the weak-$*$ topology, the $Y$-marginal of $\nu$ is $\mu$. We shall now show that $\nu$ is indeed an invariant measure satisfying \eqref{eq:measureCond}.

For the invariance of $\nu$, it is sufficient to show that for any continuous function on $X\times Y$ we have 
\[\intop f \cdot d\nu=\intop f\circ T  \cdot d \nu.\]
Indeed, 
\begin{align*}
 \intop f\cdot d\nu-\intop f\circ T \cdot d \nu& =\lim_{k\to\infty }\intop f \cdot d\nu_k-\lim_{k\to\infty }\intop  f\circ T \cdot d\nu_k \\
 &=\lim_{k\to\infty } \frac{1}{n_k}\sum_{i=0}^{n_k-1} \parenv*{\intop f \cdot d T^i_*\nu_k'-\intop f\circ T \cdot d T^i_*\nu_k'} \\
 &=\lim_{k\to\infty } \frac{1}{n_k}\sum_{i=0}^{n_k-1} \parenv*{\intop f\circ T^i \cdot d \nu_k'-\intop f\circ T^{i+1} \cdot d \nu_k'} \\
 &=\lim_{k\to\infty}\frac{1}{n_k}\parenv*{\intop f \cdot d \nu_k'-\intop f\circ T^{n_k} \cdot d \nu_k'}=0,
\end{align*}
where the convergence to $0$ follows since $f$ is bounded (as a continuous function on a compact space).

It now remains to show that $\nu$ satisfies \eqref{eq:measureCond}. From the definitions of $\nu$, $\nu_k$, and $\nu_k'$, for all $k$ we have
\begin{align*}
  \Pro_\nu[\bX_0\neq \bY_0]&=\lim_{k\to\infty}\Pro_{\nu_k}[\bX_0\neq \bY_0]=\lim_{k\to\infty}\frac{1}{n_k}\sum_{i=0}^{n_k-1} \Pro_{\nu_k'}\sparenv*{T^{-i}\set*{\bX_0\neq \bY_0}}\\
  &=\lim_{k\to\infty}\frac{1}{n_k}\sum_{i=0}^{n_k-1} \Pro_{\nu_k'}\sparenv*{\bX_i\neq \bY_i}=\lim_{k\to\infty} \E_{\nu_k'}\sparenv*{\frac{1}{n_k}\sum_{i=0}^{n_k-1}I_{\bX_i\neq\bY_i}}\\
  &=\lim_{k\to\infty} \E_{\nu_k'}\sparenv*{\frac{1}{n_k}d\parenv*{\bY_0^{n_k-1},\bX_0^{n_k-1}}}=\lim_{k\to\infty} \E_{\mu}\sparenv*{\frac{1}{n_k}d\parenv*{\bY_0^{n_k-1},\ox(\bY_0^{n_k-1})}}.
\end{align*}
We define $E_\delta$ to be the event that $d(\bY_0^{n_k-1},\ox(\bY_0^{n_k-1}))\geq n_k(R_\varepsilon(X,Y,\mu)+\delta/4)$. By \eqref{eq:CloseToEps}, for sufficiently large $k$, 
\begin{align*}
    \E_{\mu}\sparenv*{\frac{1}{n_k}d\parenv*{\bY_0^{n_k-1},\ox(\bY_0^{n_k-1})}}&\leq \E_{\mu}\sparenv*{\frac{1}{n_k}d\parenv*{\bY_0^{n_k-1},\ox(\bY_0^{n_k-1})}\cdot I_{E_\delta}}+\E_\mu[1-I_{E_\delta}]\\
    &\leq R_\varepsilon(X,Y,\mu)+\frac{\delta}{4}+(1-\Pro_\mu[E_\delta])\leq R_\varepsilon(X,Y,\mu)+\frac{\delta}{4}+\varepsilon\leq R_\varepsilon(X,Y,\mu)+\frac{\delta}{2}.
\end{align*} 
Combining the above inequality with \eqref{eq:CloseToZeroR} we conclude that
\[\Pro_\nu[\bX_0\neq \bY_0]=\lim_{k\to\infty }\Pro_{\nu_k}[\bX_0\neq \bY_0]\leq R_0(X,Y,\mu)+\frac{3\delta}{4}, \]
as desired.
\end{IEEEproof}

In the following example, we explicitly describe a sequence of extensions in $M_\cE(X,Y,\mu)$ which approximates the  essential covering radius of the $(0,k)$-RLL system from Example~\ref{ex:0kRLL} with respect to the full-shift (equipped with the uniform Bernoulli measure).

\begin{example}\label{ex:SBC0k}
Let $X_{0,k}\subseteq[2]^\Z$ denote the $(0,k)$-RLL shift as in Example~\ref{ex:0kRLL}. Let $\oy\in[2]^n$ be a finite binary word. We define $c(\oy)$ to be the length of longest zero suffix of $\oy$, formally given by
\[
c(\oy) \eqdef \max\set*{ i ; \oy=\oy_0^{n-i-1}\ozero_i }.
\]
We fix $N\in \N$ and consider the map $f^{(N)}:[2]^\Z\to X_{0,k}$ defined by
\[f^{(N)}(\by)_m=\begin{cases}
1 & c\parenv*{\by_{m-(N(k+1)-1)}^{m-1}} \equiv k \pmod{k+1}, 
\\
\by_m & \text{otherwise.}
\end{cases}\]
Clearly, $\mathrm{Im}(f)\subseteq X_{0,k}$ since no run of $k+1$ zeroes may appear in $f^{(N)}(\by)$. We note that the map $(f^{(N)},\mathrm{Id}):[2]^\Z\to X_{0,k}\times [2]^\Z $ is a sliding-block-code function (i.e., a function such that the value in each coordinate is determined by a finite block of adjacent coordinates), and therefore it is measurable and commutes with the shift transformation. 
Let $\muni$ be the uniform measure over $[2]^\Z$, and let $\nu_N$ be its pushforward measure on $X_{0,k}\times [2]^\Z$ using $f^{(N)}$. Clearly $\nu_N$ is an invariant measure, which is also ergodic (as a factor of an ergodic measure). Therefore, $\nu_N\in M_{\cE}(X_{0,k},[2]^\Z,\muni)$. We note that 
\begin{align*}
     \Pro_{\nu_N}[\bX_0\neq \bY_0]
    &=\Pro_{\muni}\sparenv*{c\parenv*{\bY_{-(N(k+1)-1)}^{-1}} \equiv k \pmod{k+1} \text{ and } \bY_0=0}\\
    &= \sum_{i=0}^{N-1} \Pro_{\muni}\sparenv*{c\parenv*{\bY_{-(N(k+1)-1)}^{-1}} =i(k+1)+k  \text{ and } \bY_0=0}
    \\
    &=\Pro_{\muni}\sparenv*{\bY_{-(N(k+1)-1)}^{0}=\ozero_{N(k+1)}}+\sum_{i=1}^{N-1} \Pro_{\muni}\sparenv*{\bY_{-i(k+1)}^0=1\ozero_{i(k+1)}}\\
    &=\frac{1}{2^{N(k+1)}}+\frac{1}{2}\sum_{i=1}^{N-1}\frac{1}{2^{{i(k+1)}}}.
\end{align*}
Taking $N\to \infty$ we obtain 
\[ \lim_{N\to\infty}\Pro_{\nu_N}[\bX_0\neq \bY_0]= \lim_{N\to\infty}\frac{1}{2^{N(k+1)}}+\frac{1}{2}\sum_{i=1}^{N-1}\frac{1}{2^{{i(k+1)}}}=\frac{1}{2(2^{k+1}-1)}=R_0(X_{0,k},[2]^\Z,\muni).\]
\end{example}

To conclude this section we briefly discuss the essential covering radius in the context of the QCC scheme. Loosely speaking, asymptotically, all but a vanishing fraction of $\Sigma^n$ may be quantized to $\sB_n(X)$ by changing an $R_0(X,\Sigma^\Z,\muni)$-fraction of the positions. This fraction may be significantly lower than the worst-case fraction $R(X,\Sigma^Z)$. In a finite-length setting, at least a $(1-\varepsilon)$-fraction of $\Sigma^n$ may be quantized to $\sB_n(X)$ by changing at most $r_\varepsilon=R_{\varepsilon}(\sB_n(X),\Sigma^n,\muni_n)$ positions. A small obstacle we need to overcome is the fact that in the QCC scheme we do not quantize any word from $\Sigma^n$, but rather only codewords of the error-correcting code $C$. The $\varepsilon$-fraction of words from $\Sigma^n$ that are a long distance from $\sB_n(X)$ may disproportionately reside in $C$. However, if we further assume that $C$ is a linear error-correcting code, by a simple averaging argument there exists at least one coset of the code, $C'$, such that the  such that the fraction of codewords whose distance to the language of $X$ is at most $r_\varepsilon$.
This means that there exists $C'' \subseteq C'$ with $|C''| \ge (1-\epsilon)|C'|$ such that $R(B_n(X),C'') \le r_\epsilon$.

\section{Upper Bounds On The Essential Covering Radius}\label{sec:CoupAlg}
The goal of this section is to establish general upper bounds on the essential covering radius. While Theorem~\ref{th:ErgdicCaracter} gives an exact expression for the essential covering radius, the minimization problem involved is hard to solve. In Example~\ref{ex:SBC0k}, we found a sequence of good extensions which approximates the essential covering radius. In general, by Theorem~\ref{th:ErgdicCaracter}, such a sequence of extensions provides a sequence of upper-bounds on the essential covering radius. In this section we shall present two different approaches for constructing extensions for general constrained systems, thus providing upper bounds on the essential covering radius. The first approach, using Markov chains, provides an upper bound which is efficiently computable as the solution of a linear-programming problem. An alternative method for constructing extensions is by sliding-block-code functions. In that case, we prove that if $X$ is primitive, the essential covering radius can be approximated by increasing the block size in such functions.

\subsection{Markov Chains}

We consider the scenario where $X$ and $Y$ are constrained systems generated by labeled graphs $G_X=(V_X,E_X,L_X)$ and $G_Y=(V_Y,E_Y,L_Y)$ respectively. Throughout this section, we assume that $G_X$ and $G_Y$ contain no parallel edges with the same label. An edge $u\to v$ shall be denoted by the ordered pair $e=(u,v)$, and we say its source is $\src(e)=u$ and its target is $\tar(e)=v$. We focus on the case where the measure $\mu\in M_\cE(Y)$ is generated by some Markov chain on the graph $G_Y$. We remark that the case of $Y=\Sigma^\Z$ and $\mu=\muni$ is the uniform Bernoulli measure, falls into that category. We begin with some definitions and basic results from the theory of Markov chains on finite graphs.

\begin{definition}
Let $G=(V,E)$ be a finite directed graph. A stationary Markov chain on $G$ is a pair $(\pi,Q)$, where $\pi$ is a probability measure on $V$ and  $Q$ is a function from $V$ to the space of probability measures on $E$ that sends $v \in V$ to a probability measure $Q(\cdot | v)$ on $E$ such that for every $v \in V$, \[\sum_{\substack{e\in E\\ \src(e)=v}} Q(e|v) =1,\]
and so that for every $v \in V$ we have:
\[ \pi(v) = \sum_{\substack{e\in E\\ \tar(e)=v}}\pi(\src(e))Q(e|\src(e)).\]

\end{definition}

Note that for any Markov chain on $(\pi,Q)$ on $G=(V,E)$, $Q(e | v) >0 $ implies that $\src(e)=v$ so we can conveniently write $Q(e)$ as an abbreviation for $Q(e| \src(e))$. In the case where $G$ is a simple graph (i.e., without parallel edges), for any edge $e=(u,v)\in E$ we use the notation $Q(v|u)$ for  $Q(e)$. Also, when $G$ is a simple graph, $Q$ may be identified with a $|V|\times |V|$ stochastic matrix (often called the transition matrix), for which $\pi$ is a left eigenvector with eigenvalue $1$.

There is a one-to-one correspondence between Markov chains on $G=(V,E)$ and probability measures on $E$ that satisfy the condition
\[\sum_{\substack{e\in E \\ \src(e)=v}}P(e)=\sum_{\substack{e\in E \\ \tar(e)=v}}P(e).\]
Indeed, such a probability measure $P$ corresponds to a stationary Markov chain $(\pi,Q)$, where 
\[\pi(v) \eqdef\sum_{\substack{e\in E\\ \src(e)=v}}P(e)=\sum_{\substack{e\in E\\ \tar(e)=v}}P(e), \]
and
\[Q(e| \src(v)) \eqdef \frac{P(e)}{\pi(\src{v})}.\]
By abuse of notation, we denote $P=(\pi,Q)$. We assume the Markov chain does not contain  degenerate vertices, i.e., $\pi(v)>0$ for all $v\in V$. Any stationary Markov chain $P$ induces an invariant measure $\hat{P}$ on the space of bi-infinite paths on $G$ by
\begin{align*}
    \hat{P}([(e_0,e_1,\dots,e_{n-1})])=P(e_0)\prod_{i=1}^{n-1} Q(e_i).
\end{align*}
for any cylinder set $[(e_0,\dots,e_{n-1})]$ corresponding to a finite path $(e_0,\dots,e_{n-1})$. We call $\hat{P}$ the stationary Markov process on $G$, induced by $P$. 
In the case where $G$ is a simple graph (i.e., without parallel edges), for any edge $e=(u,v)\in E$ we use the notation $Q(v|u)$ for  $Q(e)$.

If $G=G_Y$ generates the constrained system $Y$ by the labeling function  $L_Y$, then $P=(\pi,Q)$ induces an invariant probability measure on $Y$, which is the pushforward measure of $\hat{P}$ via the labeling function, i.e., for a cylinder set $[\oy]$,
\[\mu_P([\oy])=\sum_{\substack{\overline{e} \text{ path in }G\\
L(\overline{e})=\oy}}\pi(\src(e_0)) \prod_{i=0}^{|\oy|-1}Q(e_i).\]
We note that $\bY$ is a hidden Markov process with respect to $\mu_P$, and refer to the measure $\mu_P$ as above as \emph{the hidden Markov measure}  induced by $P$ via the labeling function $L$.

Assume that $X,Y\subseteq \Sigma^\Z$ are irreducible constrained systems given by labeled graphs $G_X$ and $G_Y$ respectively, and assume that $\mu=\mu_{P_Y}\in M_\cE(Y)$ is a measure on $Y$, induced by $P_Y$,  a stationary Markov Chain on $G_Y$. We consider the strong product graph of $G_X$ and $G_Y$ given by $G_{X\times Y}=\parenv*{V_{X\times Y},E_{X\times Y},(L_X,L_Y)}$ where $V_{X\times Y}\eqdef V_X\times V_Y$, $E_{X\times Y} \eqdef  E_X \times E_Y$ with $\sigma(e_x,e_y) = (\sigma(e_x),\sigma(e_y))$, $\tau(e_x,e_y) = (\tau(e_x),\tau(e_y))$
and  labeling function $L_{X \times Y}$ given by:
\[ L_{X\times Y}(e_x,e_y)=(L_X(e_x),L_Y(e_y)).\] 
 We note that a stationary Markov chain $P$ on $G_{X\times Y}$ naturally defines a stationary Markov process $\hat{P}$ on $G_{X\times Y}$, which induces the hidden Markov measure $\nu_{P}$ on  $X\times Y$ by the labeling function $L_{X\times Y}$. Since each edge $e\in E_{X\times Y}$ is composed of a pair $(e_x,e_y)\in E_X\times E_Y$, $\hat{P}$ may be considered as a measure on the space of pairs of bi-infinite paths where the first is a path on $G_X$ and the second is a path on $G_Y$. We denote the marginal measure of $\hat{P}$ on the space of bi-infinite paths on $G_Y$ by $(\hat{P})_Y$

For our purpose, we are interested in stationary Markov chains on $G_{X\times Y}$ such that the $Y$-marginal of the induced measure is $\mu_{P_Y}$. A sufficient condition is the following: considering $\hat{P}$ as a stationary Markov process on $G_{X\times Y}$, the $G_Y$-marginal measure of $\hat{P}$ is the Markov process $\hat{P}_Y$. We are therefore interested in the following question: given a stationary Markov chain $P$ on $G_{X\times Y}$, when is $\hat{P}$ an extension of the measure $\hat{P}_Y$?  An obvious necessary condition is that the $G_Y$-marginal of pairs in the distribution of $\hat{P}$ equals the pairs distribution of $\hat{P}_Y$. This condition may equivalently be described by
\begin{equation}
    \label{eq:MargCond1}
    P_Y(e)=\sum_{\substack{e'\in E_{X\times Y}\\ e'_y=e }}P(e'), \quad \text{for all }e\in E_Y.
\end{equation}

It is tempting to speculate that the condition given in \eqref{eq:MargCond1} is also sufficient, however it is not the case. The marginal of a Markov measure in general is not necessarily a Markov measure, but rather a hidden Markov measure. In fact, there exists an example for stationary Markov chains $P$ and $P_Y$ satisfying \eqref{eq:MargCond1} such that there is no invariant measure on the space of bi-infinite paths on $G_{X\times Y}$ which has the same pairs distribution as $\hat{P}$ and an $G_Y$-marginal that equals to the Markov process $\hat{P}_Y$. In the following lemma, we propose a sufficient condition under which the $G_Y$-marginal of $\hat{P}$ equals $\hat{P}_Y$. For ease of notation, we state and prove the claim for simple graphs. However, the claim is true without the assumption of graph simplicity, and the proof generalizes immediately from simple graphs to general labeled graphs. The corresponding condition for the general (not necessarily simple) case is given in \eqref{eq:CondMarg4}.  

\begin{lemma}\label{lem:MarkovExt}
Let $G_1=(V_1,E_1)$ and $G_2=(V_2,E_2)$ be finite simple directed graphs and let $G_1\times G_2 =G=(V,E)$ denote strong graph product of $G_1$ and $G_2$ (as defined above). For given stationary Markov chains $P=(\pi,Q)$ and $P_1=(\pi_1,Q_1)$ on $G$ and $G_1$ respectively, the $G_1$-marginal of the Markov process $\hat{P}$ is the Markov process $\hat{P_1}$ (on $G_1$) if the following conditions hold:
\begin{itemize}
    \item For all $e\in E_1$
\begin{equation}
    \label{eq:MargCond}
    P_1(e)=\sum_{\substack{e'\in E\\ e'_1=e }}P(e')
\end{equation}
\item For all $u_0\in V_2$ and $v_0,v_1\in V_1$ we have
\begin{equation}
    \label{eq:MargCond2}
    Q\parenv*{v_1|v_0,u_0}\eqdef \sum_{u_1\in V_2}Q\parenv*{(v_1,u_1)|(v_0,u_0)}=Q_1\parenv*{v_1|v_0}.
\end{equation}
\end{itemize}
\end{lemma}
\begin{IEEEproof}
The proof is a straightforward calculation. Let us denote by $(\hat{P})_1$ the marginal measure of $\hat{P}$ on $G_1$. By the law of total probability, for any path $\oee=(e_0,\dots,e_{n-1})=((v_0,v_1),(v_1,v_2),\dots,(v_{n-1},v_n))$ on $G_1$, 
\begin{align*}
    (\hat{P})_1([\oee])&=\sum_{u_0,\dots,u_{n}\in V_2}P\parenv*{(v_0,u_0),(v_1,u_1)}\prod_{i=1}^{n-1}Q\parenv*{(v_{i+1},u_{i+1})|(v_{i},u_{i})}\\
    &=\sum_{u_0,u_1\in V_2}P\parenv*{(v_0,u_0),(v_1,u_1)} \prod_{i=1}^{n-1}\parenv*{\sum_{u_{i+1}\in V_2} Q\parenv*{(v_{i+1},u_{i+1})|(v_{i},u_{i})}}\\
    &\overset{(a)}{=}\sum_{u_0,u_1\in V_2}P\parenv*{(v_0,u_0),(v_1,u_1)} \prod_{i=1}^{n-1}Q_1(v_{i+1}|v_i)\\
    &\overset{(b)}{=}P_1(v_0,v_1) \prod_{i=1}^{n-1}Q_1(v_{i+1}|v_i)=\hat{P}_1([\oee]),
\end{align*}
where (a) and (b) follow from the assumptions  \eqref{eq:MargCond} and \eqref{eq:MargCond2} respectively.
\end{IEEEproof}

We are now ready to give an upper bound on $R_0(X,Y,\mu)$, formulated as an optimization problem over stationary Markov chains. Let $P_Y$ and $P$ be stationary Markov chains  on $G_Y$ and $G_{X\times Y}$ respectively such that $P$ satisfies the conditions of Lemma~\ref{lem:MarkovExt}. Let us denote the measure on $X\times Y$ induced by $P$ by $\nu_P$, and denote the measure on $Y$ induced by $P_Y$ by $\mu$. By Lemma~\ref{lem:MarkovExt} we have $\nu_P\in M(X,Y,\mu)$ and therefore by  Theorem~\ref{th:ErgdicCaracter} we have
\[ R_0(X,Y,\mu)\leq \Pro_{\nu_P}[\bX_0\neq \bY_{0}]=\sum_{\substack{e=E_{X\times Y}\\ L_X(e)\neq L_Y(e)}}P(e).\]

We note that if we consider the Markov chain $P$ as an element in the simplex contained in $[0,1]^{E_X\times E_Y}$, we have that $\nu_{P}[\bX_0\neq \bY_{0}]$ is a linear function of $P$. In addition, we note the condition \eqref{eq:MargCond} is linear in $P$ as a constraint defined by a sum over the elements of $P$. In the case of simple graphs a straightforward calculation shows that \eqref{eq:MargCond2} is also a linear condition, since $Q\parenv{v^y_1|v^x_0,v^y_0}=Q(v^y_1|v^y_0)$ if and only if 
\begin{equation}
    \label{eq:ComdMark}\sum_{v^x_1\in V_X}P\parenv*{(v^x_1,v^y_1),(v^x_0,v^y_0)}=Q_Y(v^y_1|v^y_0)\cdot \sum_{\substack{v^x\in V_X\\v^y\in V_Y}}P\parenv*{(v^x,v^y),(v^x_0,v^y_0)}
\end{equation}
We observe that \eqref{eq:ComdMark} may be equivalently formulated as a condition on edges, under which the conclusion of Lemma~\ref{lem:MarkovExt} is true in the general case. That is, the conclusion of of Lemma~\ref{lem:MarkovExt} is true if \eqref{eq:MargCond} is satisfied and for all $e\in E_{X\times Y}$
\begin{equation}
    \label{eq:CondMarg4}
    \sum_{\substack{e'\in E_{X\times Y} \\ e'_y=e_y\\ \src(e'_x)=\src(e_x)}} P(e')=Q_Y(e_y)\sum_{\substack{e'\in E_{X\times Y} \\ \src(e'_y)=\src(e_y)\\ \src(e'_x)=\src(e_x)}}P(e')
\end{equation}

We therefore obtain an upper bound by minimizing $\Pro_{\nu_P}[\bX_0\neq \bY_0]$ over all stationary Markov chains  on $G_{X\times Y}$ satisfying \eqref{eq:MargCond} and \eqref{eq:CondMarg4}, which turns out to be a linear-programming problem.

\begin{theorem}\label{th:MarkovUP}
Let $X,Y\subseteq \Sigma^\Z$ be shift spaces defined by the labeled graphs $G_X$ and $G_Y$ respectively, and let $P_Y$ be a stationary Markov chain on $Y$ that induces the measure $\mu$. Then 
\[R_0(X,Y,\mu)\leq \MB(G_{X},G_Y,P_Y),\]
where $\MB(G_{X},G_Y,P_Y)$ is the solution to the following linear-programming problem:
\begin{mini*}|l|
{P\in \R^{{E_{X\times Y}}}}{\sum_{\substack{e\in E_{X\times Y}\\ L_X(e)\neq L_Y(e)}}P(e)}
{}{}
\addConstraint{P(e)}{\geq 0,}{\forall e\in E_{X\times Y}}
\addConstraint{\sum_{e\in E_{X\times Y}}P(e)}{=1}
\addConstraint{\sum_{\substack{e'\in E_{X\times Y}\\ e'_y=e}}P(e')}{=P_Y(e),}{\forall e\in E_Y}
\addConstraint{\sum_{\substack{e\in E_{X\times Y} \\ \src(e)=v}}P(e)}{=\sum_{\substack{e\in E_{X\times Y}\\ \tar(e)=v}}P(e),}{\forall v\in V_{X\times Y}}
\addConstraint{\sum_{\substack{e'\in E_{X\times Y} \\ e'_y=e_y\\ \src(e'_x)=\src(e_x)}} P(e')}{=Q_Y(e_y)\sum_{\substack{e'\in E_{X\times Y} \\ \src(e'_y)=\src(e_y)\\ \src(e'_x)=\src(e_x)}}P(e')\quad}{\forall e\in E_{X\times Y}.}
\end{mini*}
\end{theorem}

\begin{figure}[t]
\centering
\begin{overpic}[scale=0.35]
    {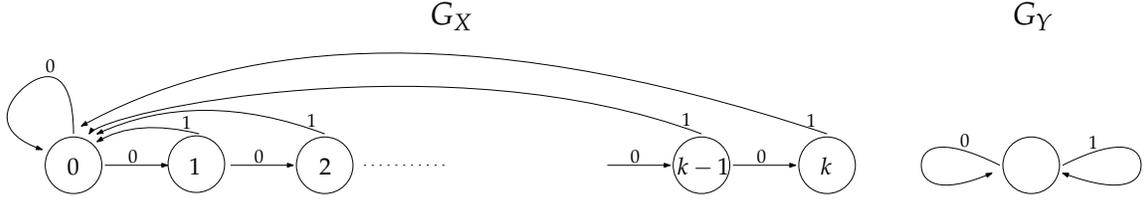}
    \put(5.4,1.8){\small{$0$}}
    \put(16.1,1.8){\small{$1$}}
    \put(27.6,1.8){\small{$2$}}
    \put(59.1,1.8){\small{$k-1$}}
    \put(71.7,1.8){\small{$k$}}
    
    \put(10.75,2.8){\scriptsize{$0$}}
    \put(21.85,2.8){\scriptsize{$0$}}
    \put(55,2.8){\scriptsize{$0$}}
    \put(66.1,2.8){\scriptsize{$0$}}
    \put(15.5,5.8){\scriptsize{$1$}}
    \put(26.5,6){\scriptsize{$1$}}
     \put(59.5,6.1){\scriptsize{$1$}}
     \put(70.5,6){\scriptsize{$1$}}
      \put(3.5,10.8){\scriptsize{$0$}}
      \put(84,4.2){\scriptsize{$0$}}
      \put(95.3,4.1){\scriptsize{$1$}}
      
      \put(37.3,15){\large{$G_X$}}
      \put(88.6,15){\large{$G_Y$}}
\end{overpic}
\caption{
Labeled graphs generating the shift spaces $X=X_{0,k}$ (the $(0,k)$-RLL shift), and $Y=[2]^\Z$.}
\label{fig:0K-UN}
\end{figure}

\begin{figure}[t]
\centering
\begin{overpic}[scale=0.22]
    {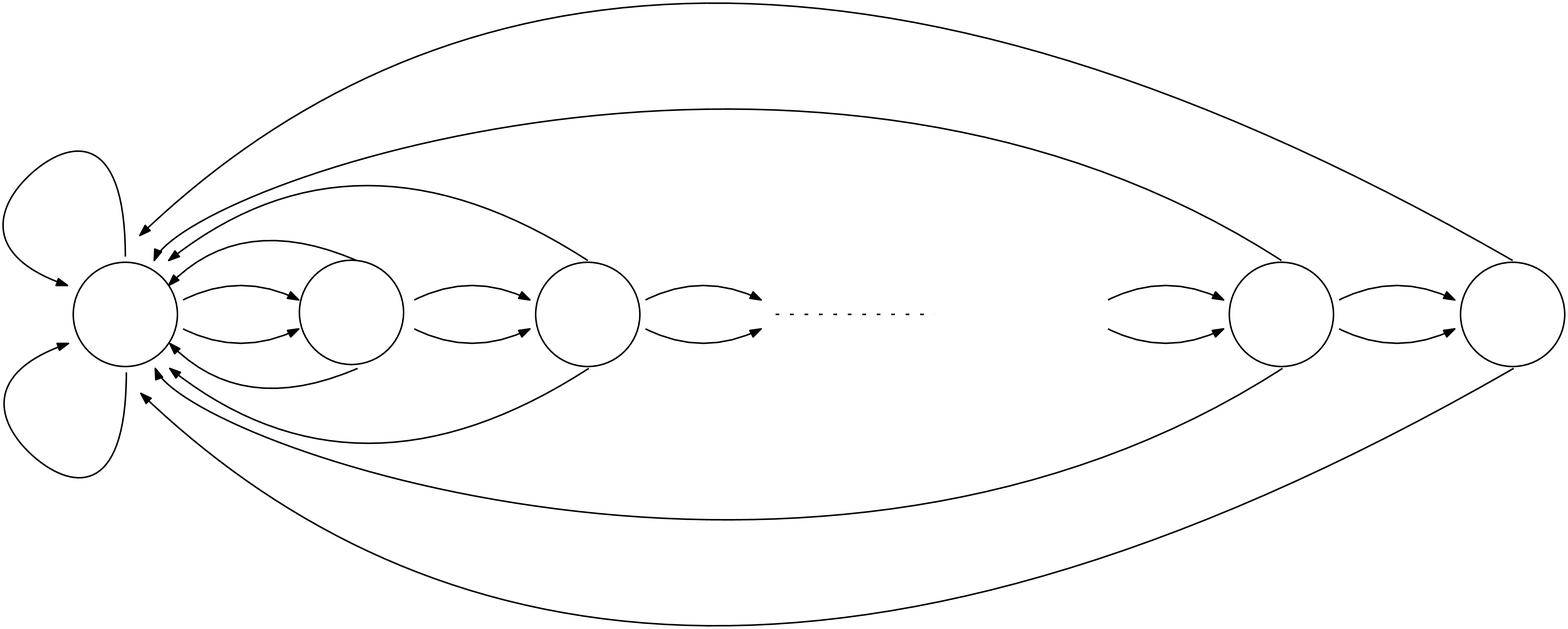}
    \put(7.2,19.1){\small{$0$}}
    \put(21.8,19.1){\small{$1$}}
    \put(36.8,19.1){\small{$2$}}
    \put(79,19.1){\small{$k-1$}}
    \put(96,19.1){\small{$k$}}
    
    \put(13,22){\scriptsize{$00,\textcolor{blue}{2^k \alpha}$}}
    \put(26.5,22){\scriptsize{$00,\textcolor{blue}{2^{k-1} \alpha}$}}
      \put(41,22){\scriptsize{$00,\textcolor{blue}{2^{k-2} \alpha}$}}
      \put(71.7,22){\scriptsize{$00,\textcolor{blue}{4\alpha}$}}
    \put(86,22){\scriptsize{$00,\textcolor{blue}{2\alpha}$}}
    \put(16,25){\scriptsize{$10,\textcolor{red}{0}$}}
    \put(34,25.5){\scriptsize{$10,\textcolor{red}{0}$}}
      \put(79.3,25){\scriptsize{$10,\textcolor{red}{0}$}} \put(93.9,25){\scriptsize{$10,\textcolor{blue}{\alpha}$}} \put(2,31){\scriptsize{$00,\textcolor{blue}{2^k \alpha}$}}

        \put(14,16.5){\scriptsize{$01,\textcolor{red}{0}$}}
        \put(27.5,16.5){\scriptsize{$01,\textcolor{red}{0}$}}
        \put(42,16.5){\scriptsize{$01,\textcolor{red}{0}$}}
        \put(72.7,16.5){\scriptsize{$01,\textcolor{red}{0}$}}
        \put(87,16.5){\scriptsize{$01,\textcolor{red}{0}$}}
        \put(15.5,13.5){\scriptsize{$11,\textcolor{blue}{2^{k-1}\alpha}$}}
        \put(34.5,13.3){\scriptsize{$11,\textcolor{blue}{2^{k-2}\alpha}$}}
        \put(78.9,13.3){\scriptsize{$11,\textcolor{blue}{2\alpha}$}}
        \put(94,13.3){\scriptsize{$11,\textcolor{blue}{\alpha}$}}
        \put(43.5,36){\large{$G_{X\times Y}$}}
        \put(2.9,7.7){\scriptsize{$01,\textcolor{red}{0}$}}
\end{overpic}
\caption{
The product graph $G_{X\times Y}$ for the graphs $G_X$ and $G_Y$ from Figure~\ref{fig:0K-UN}. Each edge is given a two-bit label, $xy$, corresponding to the label $x$ from $G_X$ and the label $y$ from $G_Y$. A stationary Markov chain achieving the bound $MB(G_X,G_Y,P_Y)$ is shown by writing $P(e)$ after the label on each edge. Edges with $P(e)=0$ are marked in red, and positive probabilities $P(e)>0$ are written in blue. }
\label{fig:0Kprod}
\end{figure}

\begin{example}
Consider $X_{0,k}$, the $(0,k)$-RLL shift from Example~\ref{ex:0kRLL}. Take $Y=[2]^\Z$ and let $\muni$ be the uniform Bernoulli measure on $Y$. We consider the labeled graphs $G_X$ and $G_Y$ shown in Figure~\ref{fig:0K-UN}, generating $X$ and $Y$ respectively. We take $P_Y$ to be the uniform measure on $E_Y$ (inducing the measure $\muni$). The product graph, $G_{X\times Y}$ (shown in Figure~\ref{fig:0Kprod}), is therefore a ``doubled'' version of the graph $G_X$.

We consider the Markov measure $P$, defined by the edge probabilities given in Figure~\ref{fig:0Kprod}. For an appropriate choice of $\alpha$, $P$ is indeed a stationary Markov chain satisfying \eqref{eq:MargCond} and \eqref{eq:CondMarg4}. First, in order to get a probability measure on edges we require
\[ 1=\sum_{e\in G_{X\times Y}}P(e)=2\alpha(2^k+2^{k-1}+\cdots + 2+1)=2\alpha(2^{k+1}-1),\]
which implies that $\alpha=\frac{1}{2(2^{k+1}-1)}$. We observe that for the $j$-th state in $V_{X\times Y}$, 
\[\sum_{\src(e)=j}P(e)=\sum_{\tar(e)=j}P(e)=\alpha\cdot 2^{k-j},\]
which implies that $P$ is indeed stationary. We also observe that for any edge with $L_Y(e)=0$ there is a corresponding edge $e'$ with $L_Y(e')=1$ such that $P(e)=P(e')$. This shows that the marginal of $P$ on $G_Y$ is indeed $P_Y=\muni$.  We further observe that for any edge $e\in E_{X\times Y}$, if $\src(e_x)=j\in V_X$  we have
\[\sum_{\substack{e'\in E_{X\times Y} \\ e'_y=e_y\\ \src(e'_x)=j}} P(e')=\alpha 2^{k-j-1}=\frac{1}{2}\alpha 2^{k-j}=Q_Y(e_y)\sum_{\substack{e'\in E_{X\times Y} \\ \src(e'_y)=\src(e_y)\\ \src(e'_x)=j}}P(e') .\]

We now compute: 
\[\sum_{\substack{ e\in G_{X\times Y}\\ L_X(e)\neq L_Y(e)}}P(e)=\alpha=\frac{1}{2(2^{k+1}-1)}=R_0(X_{0,k},[2]^\Z,\muni).\]
Thus, by Theorem~\ref{th:0k} and Theorem~\ref{th:MarkovUP}, $P$ attains the minimal value for the linear-programming problem $\MB(G_X,G_Y,P_Y)$, and in particular in this case, the upper bound from Theorem~\ref{th:MarkovUP} is tight.
\end{example}

\begin{figure}[t] 
\centering\scalebox{1.15}{
\begin{overpic}[scale=0.27]
    {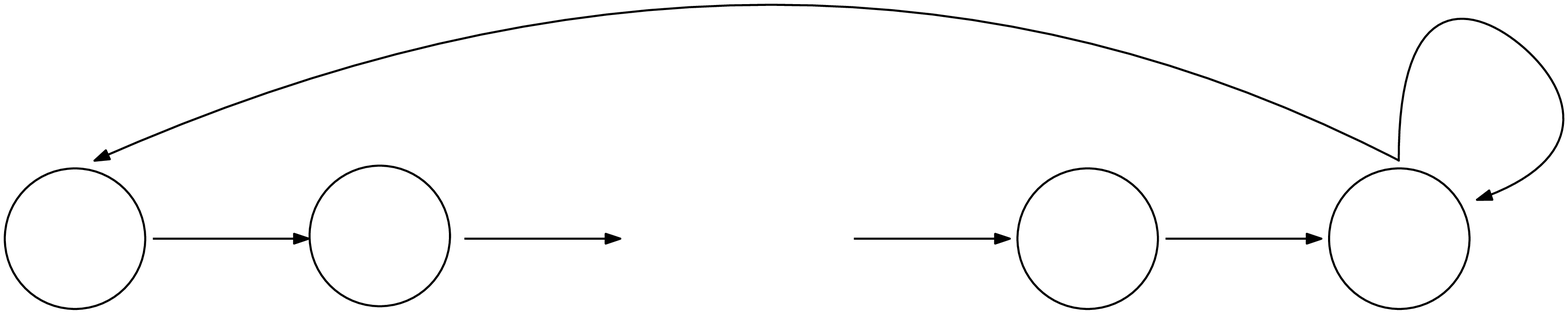}
    \put(3.83,3.5){\small{$0$}}
    \put(23.3,3.3){\small{$1$}}
    \put(65.8,3.3){\small{$d-1$}}
    \put(88.5,3.3){\small{$d$}}
    
    \put(13.6,5){\small{$0$}}
    \put(33.3,5){\small{$0$}}    
    \put(58.3,5){\small{$0$}}
    \put(78.3,5){\small{$0$}}    
    \put(78.3,15.1){\small{$1$}}    
    \put(95.3,12.1){\small{$0$}}    
       \put(43,22){ \large{$G_{X_{d,\infty}}$}}   
\end{overpic}}
\caption{
A labeled graph generating the constrained system $X=X_{d,\infty}$ (the $(d,\infty)$-RLL shift).}\label{fig:Dinfty-UN}
\end{figure}

\begin{figure}[t]
\centering\scalebox{1.15}{
\begin{overpic}[scale=0.270]
    {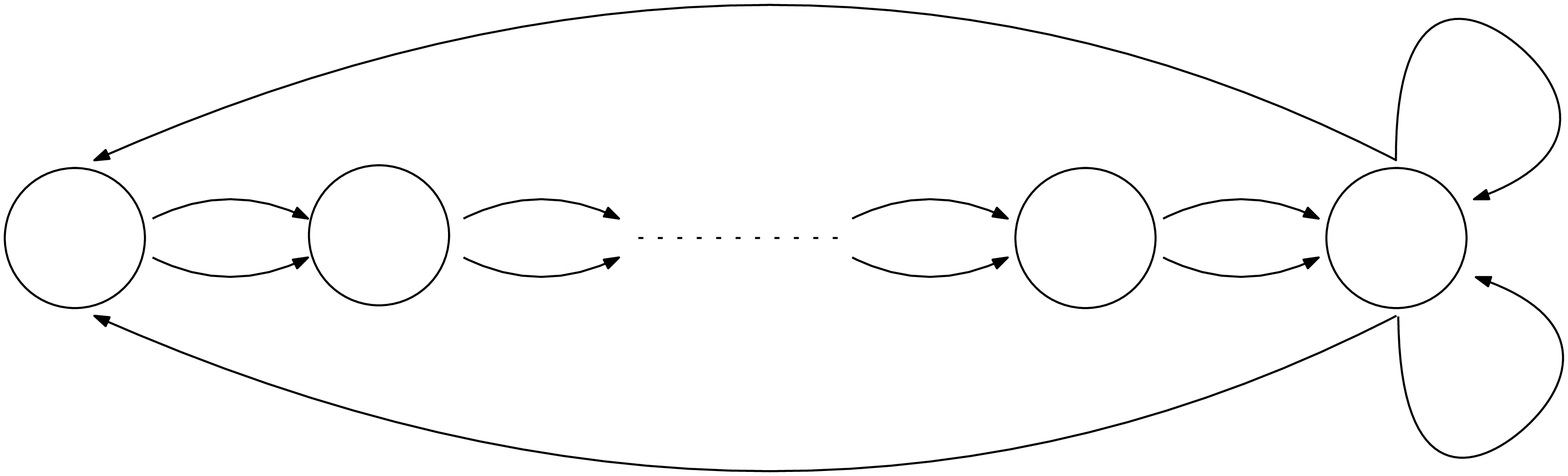}
    \put(3.79,13.7){\small{$0$}}
    \put(23.3,13.7){\small{$1$}}
    \put(65.8,13.7){\small{$d-1$}}
    \put(88.5,13.7){\small{$d$}}
    
    \put(11.8,17.9){\scriptsize{$00,\textcolor{blue}{\alpha}$}}
    \put(32.4,17.9){\scriptsize{$00,\textcolor{blue}{\alpha}$}}    
    \put(57,17.9){\scriptsize{$00,\textcolor{blue}{\alpha}$}}    
    \put(77,17.9){\scriptsize{$00,\textcolor{blue}{\alpha}$}} 
    \put(77,26){\scriptsize{$10,\textcolor{red}{0}$}}  
    \put(93.4,21.9){\scriptsize{$00,\textcolor{blue}{\alpha}$}}   
    
    \put(11.8,10.5){\scriptsize{$01,\textcolor{blue}{\alpha}$}}
    \put(32.4,10.5){\scriptsize{$01,\textcolor{blue}{\alpha}$}}    
    \put(57,10.5){\scriptsize{$01,\textcolor{blue}{\alpha}$}}    
    \put(77,10.5){\scriptsize{$01,\textcolor{blue}{\alpha}$}} 
    \put(77,2.5){\scriptsize{$11,\textcolor{blue}{\alpha}$}}  
    \put(93.4,6.5){\scriptsize{$01,\textcolor{red}{0}$}}   
   \put(43,25){ \large{$G_{X_{d,\infty}\times Y}$}}   
\end{overpic}}
\caption{
The product graph $G_{X\times Y}$ for the graphs $G_{X_{d,\infty}}$ and $G_Y$ from Figure~\ref{fig:Dinfty-UN} and Figure~\ref{fig:0K-UN} respectively. Each edge is given a two-bit label, $xy$, corresponding to the label $x$ from $G_{X_{d,\infty}}$ and the label $y$ from $G_Y$. A stationary Markov chain achieving the bound $MB(G_X,G_Y,P_Y)$ is shown by writing $P(e)$ after the label on each edge. Edges with $P(e)=0$ are marked in red, and positive probabilities $P(e)>0$ are written in blue.
}
\label{fig:Dinfty-Prod}\end{figure}

\begin{example}\label{ex:Dinfty}
Let $X=X_{d,\infty}$ be the $(d,\infty)$-RLL system, defined by the constraint of having a run of at least $d$ zeroes between any two consecutive ones.  Equivalently, $X_{d,\infty}$ is defined by $G_X$ presented in Figure~\ref{fig:Dinfty-UN}. Let $Y=[2]^\Z$ and $\muni$ be as in Example~\ref{ex:0kRLL}.  The product graph $G_{X\times Y} $ is shown in Figure~\ref{fig:Dinfty-Prod}. We consider the Markov measure $P$, defined by the edge probabilities given in Figure~\ref{fig:Dinfty-Prod}. For  $\alpha=(2(d+2))^{-1}$ we have 
\begin{align*}
    1&=\sum_{e\in G_{X\times Y}}P(e)=2\alpha(d+2),
\end{align*} 
and so $P$ is indeed a stationary Markov chain satisfying the conditions of Theorem~\ref{th:MarkovUP}.
We now compute the upper-bound: 
\begin{equation}
    \label{eq:dinftyB}
    R_0(X_{d,\infty},[2]^\Z,\muni)\leq \sum_{\substack{ e\in G_{X\times Y}\\ L_X(e)\neq L_Y(e)}}\hspace{-3ex}P(e)=\alpha\cdot d=\frac{d}{2(d+2)}.
\end{equation}
We note that when $d=1$, the system $X_{1,\infty}$ is isomorphic to  $X_{0,1}$, by complementing all the bits. Thus 
\[R_0(X_{1,\infty},[2]^\Z,\muni)=R_0(X_{0,1},[2]^\Z,\muni)=\frac{1}{6},\]  and in particular, the bound \eqref{eq:dinftyB} is tight.

For a lower bound, we  claim that \[R_0(X_{d,\infty},[2]^\Z,\muni)\geq \frac{1}{2}-\frac{1}{d+1}.\]
The proof is straight-forward from the definition of the essential covering radius. We note that for $d=1$ the bound is meaningless, and therefore we assume that $d\geq 2$. We start by a simple observation. Let $\wt(\ow)=d(\ow,\ozero)$ denote the number of non-zero coordinates of a binary word $\ow$. We note that for any $\ox,\oy\in [2]^n$,
\[d(\ox,\oy)\geq \abs{\wt(\oy)-\wt(\ox)} \geq \wt(\oy)-\wt(\ox). \]
From the definition of the $X_{d,\infty}$ system, for any any $\ox\in \sB_n(X)$, there must be at least $d$ zeroes between any two consecutive ones. Hence, for any $\ox\in \sB_n(X)$ we have  
\[\wt(\ox)\leq \ceil*{\frac{n}{d+1}}\leq \frac{n}{d+1}+1.\]
Thus, for any word $\oy\in [2]^n$, we have 
\begin{equation}\label{eq:LBdINFTY}
    \min_{\ox\in \sB_n(X)}d(\ox,\oy)\geq \wt(\oy)- \frac{n}{d+1}-1.
\end{equation}

Let $\bY$ be a random bi-infinite sequence generated by the distribution $\muni$. That is, $(\bY_k)_{k\in\Z}$ are i.i.d $\mathrm{Ber}(\frac{1}{2})$ random variables. Applying the law of large numbers, we have 
\[\frac{1}{n}\wt\parenv*{\bY_0^{n-1}}=\frac{1}{n}\sum_{i=0}^{n-1}\bY_i\xrightarrow[n\to \infty]{\Pro_{\muni}}\E_{\muni}[\bY_0]=\frac{1}{2}.\]
Namely, the normalized weights of random words converge in $\Pro_{\muni}$ probability to $\frac{1}{2}$. Combining this with \eqref{eq:LBdINFTY}, we obtain that for every $\delta>0$
\begin{align*}
    \Pro_{\muni_n}&\sparenv*{\min_{\ox\in \sB_n(X_{d,\infty})}\frac{d(\ox,\bY_0^{n-1})}{n}>\frac{1}{2}-\frac{1}{d+1}-\frac{1}{n}-\delta}\\
    &\geq \Pro_{\muni} \sparenv*{\frac{1}{n}\wt(\bY_0^{n-1})>\frac{1}{2}-\delta}\xrightarrow[n\to\infty]{}1.
\end{align*}
This proves that for any $\varepsilon\in (0,1)$, $\delta>0$, for sufficiently large $n$
\[\frac{R_{\varepsilon}(\sB_n(X_{d,\infty}),\sB_n(Y),\muni)}{n}> \frac{1}{2}-\frac{1}{d+1}-\delta,\]
and therefore, 
\[R_\varepsilon(X_{d,\infty},[2]^\Z,\muni)\geq \lim_{\delta \to 0}\frac{1}{2}-\frac{1}{d+1}-\delta =\frac{1}{2}-\frac{1}{d+1}.\]
Taking $\varepsilon\to 0$, we get
\[R_0(X_{d,\infty},[2]^\Z,\muni)\geq \frac{1}{2}-\frac{1}{d+1}.\]
Combining the lower and upper bounds we get,
\[\frac{1}{2}-\frac{1}{d+1}\leq R_0(X_{d,\infty},[2]^\Z,\muni)\leq \frac{1}{2}-\frac{1}{d+2}.\]
\end{example}

\subsection{Sliding Block Codes}

We now present an alternative approach for constructing extensions by using sliding-block-codes functions. We begin by revising Example~\ref{ex:SBC0k}, where we found a sequence of extensions approximating the essential covering radius of $X_{0,k}$ with respect to $[2]^\Z$ with $\muni$, (the uniform i.i.d measure). The main idea in the construction of these extensions was the following: for $X,Y\subseteq \Sigma^\Z$ and $\mu\in M_\cE(Y)$, given a measurable function $g:Y\to X$ which commutes with the shift transformation, the map $(g,\mathrm{Id}):Y\to X\times Y$ defines an extension $\nu_g$ in $M(X,Y,\mu)$ by the pushforward of $\mu$ via $(g,\mathrm{Id})$. That is 
\[ \nu_g(A_X\times A_Y)\eqdef \mu\parenv*{A_Y\cap g^{-1}(A_X)}. \]
We call such a function $g$ a \emph{stationary coding function} from $Y$ to $X$. By Proposition~\ref{prop:JoinBound}, for any such stationary coding function $g$,
\[R_0(X,Y,\mu)\leq \Pro_{\nu_g}[\bX_0\neq \bY_0],\]
Namely, any such measurable function $g$ that commutes with the shift provides an upper bound on the essential covering radius. 

In Example~\ref{ex:SBC0k}, we used sliding-block-code functions (functions defined by a local rule) as our measurable functions that commute with the shift. Sliding-block-code functions are of particular interest to us since they provide a rich family of functions, easily described by a local rule.  The properties and constructions of sliding-block codes have been extensively studied in the literature (for example, see~\cite{ashley1995canonical, ashley1996complexity, adler1983algorithms}). The goal of this section is to explicitly describe the bound obtained from a sliding-block-code function, and to give sufficient conditions under which the essential covering radius can be approximated using extensions constructed by sliding-block codes. 

\begin{definition}
Let $X, Y\subseteq \Sigma^\Z$ be shift spaces. A function $\hat{f}:Y\to X$ is called a \emph{sliding-block code} if there exist $N\in \N$ and a function $f:\sB_{2N+1}(Y)\to \Sigma$ such that for all $\by\in Y$ and all $i\in\Z$,
\[\hat{f}(\by)_i=f(\by_{i-N}^{i+N}).\]
In that case, $\hat{f}$ is said to be a sliding-block code of block length $N$.
\end{definition}

Let $\hat{f}:Y\to X$ be a sliding-block-code function defined by a local function $f:\sB_{2N+1}(Y)\to \Sigma$, and let $\mu\in M_\cE(Y)$ be an ergodic measure. We denote the extension obtained from $\hat{f}$ by $\nu_f$. The quantity $\Pro_{\nu_f}[\bX_0\neq\bY_0]$ is now easily computable: 
\begin{align*}
    \Pro_{\nu_f}[\bX_0\neq\bY_0]=\Pro_{\mu}[\hat{f}(\bY)_0\neq\bY_0]=\Pro_{\mu}\sparenv*{f(\bY_{-N}^{N})\neq\bY_0}=\sum_{\substack{\oy\in \sB_{2N+1}(Y)\\
    f(\oy)\neq \oy_N}}\mu([\oy]). 
\end{align*}
We now state the main result of this section: if $X$ is a primitive constrained system and $\mu$ is aperiodic (namely, the $\mu$-measure of all shift-periodic points of $Y$ is zero) then the essential covering radius $R_0(X,Y,\mu)$ may be approximated by extensions constructed by sliding-block codes, as the block length is increased. 

\begin{theorem}\label{th:SBCchar}
Let $X,Y\subseteq \Sigma^\Z$ be constrained systems such that $X$ is primitive and $\mu\in M_\cE(Y)$ is an aperiodic ergodic measure. Then for any $\varepsilon>0$ there exists a sufficiently large $N$ and a sliding-block-code function $\hat{f}$ of length $N$ such that 
\[ \Pro_{\nu_f}[\bX_0\neq\bY_0] -\varepsilon \leq R_0(X,Y,\mu)\leq  \Pro_{\nu_f}[\bX_0\neq\bY_0]. \] 
\end{theorem} 

The proof of Theorem~\ref{th:SBCchar} has two major components. The first part is showing that it is possible to approximate the essential covering radius by extensions obtained from stationary coding functions. In the second step, we use the fact that sliding-block-code functions are dense in the space of stationary coding functions to obtain the main result.

For the first component of the proof, we shall require an improved version of Alpern's Lemma~\cite{alpern1979generic}, recently proved  in~\cite{campbell2019independence}.

\begin{definition}
An ergodic measure $\mu\in M_{\cE}(Y)$ is called \emph{aperiodic} if the measure of periodic points is $0$. Namely,
\[\mu\parenv*{\set*{\by\in Y; \exists n\in \N \text{ such that }T^n\by=\by}}=0.\]
\end{definition}

\begin{lemma}[{{\cite[Theorem 1]{campbell2019independence}}}]
\label{lem:Alp}
Let $Y\subseteq \Sigma^\Z$ be a constrained system, $\mu\in M_{\cE}(Y)$ be an aperiodic ergodic measure, $n_1,\dots,n_k$ be pairwise co-prime integers and $q_1\dots,q_k$ positive numbers such that  $\sum_{i=1}^k q_in_i=1$. For any finite measurable partition $\cP$ of $Y$ there exist measurable sets $Q_1,\dots, Q_k$ such that: 
\begin{enumerate}
    \item The set 
\[\cP'=\set*{T^{-i}(Q_j); 1\leq j\leq k , 0\leq i \leq n_j-1}\]
is a partition of $Y$.
\item For all $1\leq j\leq k$, $\mu(Q_j)=q_j$.
\item For all $1\leq j\leq k$, $Q_j$ is independent of the partition $\cP$ with respect to $\mu$ (namely, for any $A\in \cP$ $\mu(A\cap Q_j)=\mu(A)\mu(Q_j)$).
\end{enumerate}
\end{lemma}  

\begin{theorem}
\label{th:JoinTight}
Let $X,Y\subseteq \Sigma^\Z$ be constrained systems, $X$ be primitive, and $\mu\in M_{\cE}(Y)$ be an aperiodic ergodic measure. Then 
\[R_0(X,Y,\mu)=\inf\set*{\Pro_{\nu_g}[\bX_0\neq \bY_0]; g:Y\to X \text{ is a stationary coding function}}.\] 
\end{theorem}

\begin{IEEEproof}
The proof of this theorem is quite involved, and so we would like to start by outlining the proof strategy. By Theorem~\ref{th:ErgdicCaracter}, it is sufficient to find a sequence of extensions obtained from stationary coding functions such that the probability of the event $\set{\bX_0\neq \bY_0}$ converges to $R_0(X,Y,\mu)$. We construct these stationary coding functions maps by block coding. Using Alpern's Lemma, we divide an infinite sequence from $Y$ into blocks of two types, one large and the other small. The large blocks are common while the small occur only rarely. The large blocks we code (with high probability) to a close counterpart from the language of $X$, such that the whole bi-infinite sequence ends in $X$. This is possible since $X$ is primitive. We show that using this kind of block coding, the probability of the event $\set{\bX_0\neq \bY_0}$ converges to $R_0(X,Y)$.

We now start with the proof. Let $X$ be defined by the primitive labeled graph $G=(V,E,L)$. We fix an arbitrarily small $\varepsilon\in (0,1)$. Since $G$ is primitive, there exists a number $p$ such that for any two vertices $v_1,v_2\in V$, there exists a directed path of length $p$ from $v_1$ to $v_2$. For any $v_1,v_2\in V$ we fix such a directed path of length $p$, which we denote by $\Gamma(v_1,v_2)$. Given any directed path on the graph $\Gamma=e_1,\dots,e_m$ we define $\src(\Gamma)\eqdef\src(e_1)$ and $\tar(\Gamma)\eqdef\tar(e_m)$. We also define $L(\Gamma)\eqdef (L(e_1),\dots, L(e_m))\in \sB_m(X)$ to be the word obtained by reading the labels on $\Gamma$.

In order to simplify the notation, for any $M\in \N$ we denote $r_\varepsilon(M)\eqdef R_\varepsilon(\sB_M(X),\sB_M(Y),\mu_M)$. By the definition of $R_{\varepsilon}(X,Y,\mu)$, there exists a sequence of positive integers $(M_i)_{i\in\N}$, $M_i\to \infty$, and a vanishing sequence of numbers $(\delta_i')_{i\in\N}$ 
\[\frac{r_{\varepsilon}(M_i)}{M_i}= R_{\varepsilon}(X,Y,\mu)+\delta_i',\]
for all $i\in\N$. We let $N_i\eqdef \ceil{\frac{M_i}{p}}$, and we easily observe that 
\[ r_{\varepsilon}(N_ip+1)\leq r_\varepsilon(M_i)+p.\]
Thus, 
\begin{align*}
    \frac{r_{\varepsilon}(N_i p+1)}{N_ip+1}&\leq  \frac{ r_{\varepsilon}(M_i)+p}{N_i p+1}\\
    &=\frac{M_i}{N_i p+1}\cdot \frac{r_{\varepsilon}(M_i)}{M_i} +\frac{p}{N_i p+1}\\
    &=R_{\varepsilon}(X,Y,\mu)+\delta_i,
\end{align*}
where $(\delta_i)_{i\in\N}$ is a vanishing sequence. 

Fix any $i\in \N$. From the definition of $r_\varepsilon(n)$, there exist a map $\varphi_i:\sB_{N_ip+1}(Y)\to E^{N_ip+1}$ and a set $S_{i}\subseteq \sB_{N_ip+1}(Y)$ of $\mu_{N_ip+1}$-measure at least $1-\varepsilon$,  such that for all $\oy\in S_i$, $\varphi_i(\oy)$ is a directed path on $G$ and 
\begin{equation}
\label{eq:goodword}
d\parenv*{\oy,L(\varphi_i(\oy))}\leq  r_\varepsilon(N_ip+1) \leq (N_ip+1)\cdot \parenv*{R_\varepsilon(X,Y,\mu)+\delta_i}.
\end{equation}
We call the words in $S_i$ \emph{good words}.

We recall that the measure $\mu$ is an aperiodic and ergodic,  and we note that the positive integers $n_1=p$ and $n_2=(N_i+1)p+1$ are co-prime. Defining $q_1=\frac{1}{p \cdot N_i}$ and $q_2=(1-\frac{1}{N_i})\frac{1}{(N_i+1)p+1}$, we clearly have $q_1  n_1+q_2 n_2=1$.
We consider the partition defined by the $(N_i+1)p+1$ first coordinates of $\bY$. That is
\[\cP^{(i)}\eqdef\set*{[\oy] ;\oy\in \sB_{(N_i+1)p+1 }(Y)},\]
where we recall the definition of a cylinder from~\eqref{def:cylinder}. By Lemma~\ref{lem:Alp} there exist measurable sets $Q_1,Q_2\subseteq Y$ such that the set 
\[\cP_i =\set*{T^{-\ell}(Q_j); 0\leq \ell \leq n_j-1, \quad j=1,2 }\]
is a partition of $Y$, as well as for $j=1,2$ we have $\mu(Q_j)=q_j$ and $Q_j$ is independent of $\cP^{(i)}$. We shall use the partition $\cP_i$ in order to divide a bi-infinite sequence to blocks.

For a fixed $\by\in Y$, and $m\in \Z$, we say that $\by$ admits a block of length $n_j$, $j=1,2$, in the coordinates $m,m+1,\dots, m+n_j-1$ if $T^m(\by)\in Q_j$. We enumerate the blocks composing $\by$ and denote them by $\parenv*{B_k(\by)=(m_k(\by),l_k(\by))}_{k\in \Z}$, where $B_0(\by)$ is the block containing the $0$ coordinate, $m_k$ is the starting point of the $k$-th block and $l_{k}\in \set{n_1,n_2}$ is its length. By the construction of the partition $\cP_i$, each and every coordinate in $\Z$ belongs to exactly one block. Therefore, in order to define a function $Y\to X$ it is sufficient to define the values that it takes in each and every block.

Let us fix some vertex $v_0\in V$. Given a $\by\in Y$ and the corresponding  sequence of blocks $\parenv*{B_k(\by)=(m_k(\by),l_k(\by))}_{k\in \Z}$ we define $f_i$ as follows:
\begin{itemize}
    \item We first define $f_i$ on blocks of length $n_1=p$ followed by another block of the same size to be the labels on the self loop $\Gamma(v_0,v_0)$. 
    \item We then define the values on blocks of length $n_2$: for such a block, we apply the transformation $\varphi_i$ on the first $N_ip+1$ coordinates and read the labels on the edges in the corresponding coordinates. For the remaining $p$ coordinates of the block we have two options: 
    \begin{enumerate}
        \item If the next block is of length $n_1=p$, we read the labels on the path $\Gamma(\tilde{v},v_0)$ where $\tilde{v}$ is the last vertex in the path corresponding to the first $N_ip+1$ coordinates of the block.
        \item If the next block is of length $n_2=(N_i+1)p+1$, we read the labels on the path $\Gamma(\tilde{v},v')$ where $v'$ is the initial vertex in the path corresponding to the $N_ip+1$ first coordinates of the next block.
    \end{enumerate}
    \item It remains to define the values on blocks of length $n_1$ followed by blocks of length $n_2$. In this case, we read the labels on the path $\Gamma(v_0,v')$, where $v'$ is as in the previous case.
\end{itemize}
In summary: 
\[f_i(\by)_{m_k}^{m_k+l_{k}-1}= \begin{cases}
L(\Gamma(v_0,v_0)) & l_{k}=l_{k+1}=n_1,\\
L\parenv*{\Gamma\parenv*{v_0,\src\parenv*{\varphi_i(\by_{m_{k+1}}^{m_{k+1}+N_ip}}}} & l_{k}=n_1,l_{k+1}=n_2,\\
L\parenv*{\varphi_i(\by_{m_{k}}^{m_{k}+N_ip})}L\parenv*{\Gamma\parenv*{\tar\parenv*{\varphi_i(\by_{m_{k}}^{m_{k}+N_ip})},v_0}} & l_{k}=n_2,l_{k+1}=n_1,\\
L\parenv*{\varphi_i(\by_{m_{k}}^{m_{k}+N_ip})}L\parenv*{\Gamma\parenv*{\tar\parenv*{\varphi_i(\by_{m_{k}}^{m_{k}+N_ip})},\src\parenv*{\varphi_i(\by_{m_{k+1}}^{m_{k+1}+N_ip})}}} & l_{k}=l_{k+1}=n_2.
\end{cases}\]

By the construction of $f_i$, it is easy to check that that for all $\by$, $f_i(\by)$ corresponds to a labeled bi-infinite path in $G$, and therefore $\mathrm{Im}(f_i)\subseteq X$.  We observe that $f_i$ is also measurable as $\cP_i$ is a measurable  partition. We note that for any $\by\in Y$, from the definition of the block partition, the block partition of $T(\by)$ is a shifted version of the block partition of $\by$. Thus, $f_i$ also commutes with the shift, which makes it a stationary coding function. Let $\nu_{f_i}$ be the pushforward measure of $\mu$ with respect to the transformation of $(f_i,\mathrm{Id})$.

We now turn to prove that $\Pro_{\nu_{f_i}}[\bX_0\neq \bY_0]$ indeed approximates $R_0(X,Y,\mu)$. Let $\bX$ and $\bY$ denote the random bi-infinite sequences in $\Sigma^\Z$ generated with respect to $\nu_{f_i}$, and let $I_{\set{\bX_0\neq\bY_0}}$ be the indicator function of the event $\set{\bX_0\neq \bY_0}$. Since $\nu_{f_i}$ is an invariant measure, for all $n\in \N$
\[\Pro_{\nu_{f_i}}[\bX_0\neq \bY_0]=\intop I_{\set{\bX_0\neq\bY_0}}\cdot d\nu_{f_i} = \intop \frac{1}{n}\sum_{k=0}^{n-1} I_{\set{\bX_0\neq\bY_0}}\circ T^{k} \cdot d\nu_{f_i}=\E_{\nu_{f_i}}\sparenv*{\frac{1}{n}d\parenv*{\bX_0^{n-1},\bY_0^{n-1}}}. \]

Let $C_n$ be the random variable that counts the number of good blocks in coordinates $0,1,\dots, n-1$. These are blocks of length $n_2=(N_i+1)p+1$ contained in $\bY_{0}^{n-1}$ with the first $N_ip+1$ coordinates containing a good word, i.e., a word in the set $S_i$. Formally, for a sequence $\by\in Y$, with corresponding blocks $(B_k(\by)=(m_k(\by),l_k(\by))_{k\in\Z}$
\[C_n(\by)\eqdef\abs*{\set*{k\in \Z ; l_k=n_2,  [m_k,m_k+n_2-1]\subseteq [0,n-1]\text{ and } \by_{m_k}^{m_k+N_ip}\in S_i}}  .\]

By the construction of $f_i,\varphi_i$, $S_i$, and by~\eqref{eq:goodword}, the number of coordinates inside a single good block in which $\bX$ and $\bY$ do not agree is upper bounded by $(N_ip+1)(R_\varepsilon(X,Y,\mu)+\delta_i) + p$. Thus, we have
\begin{align*}
    d\parenv*{\bX_0^{n-1},\bY_0^{n-1}}
    &\leq C_n(\bY_0^{n-1}) \cdot ((N_ip+1)(R_\varepsilon(X,Y,\mu)+\delta_i) + p)+n-C_n(\bY_0^{n-1})((N_i+1)p+1) \\
    &= C_n(\bY_0^{n-1}) \cdot (N_ip+1)(R_\varepsilon(X,Y,\mu)+\delta_i - 1)+n
\end{align*}
Thus, 
\begin{equation}
    \label{eq:EqualityExp}
\E_{\nu_{f_i}}\sparenv*{d\parenv*{\bX_0^{n-1},\bY_0^{n-1}}}\leq \E_{\nu_{f_i}}[C_n(\bY_0^{n-1})] \cdot (N_ip+1)(R_\varepsilon(X,Y,\mu)+\delta_i-1)+n.
\end{equation}

Let $A$ be the event that a good block starts at the $0$ coordinate. That is, 
\[A=Q_2\cap\set*{\bY_{0}^{N_ip}\in S_i}.\]
We note that we can write $C_n(\bY_0^{n-1}) = \sum_{k=0}^{n-n_2}I_{A} \circ T^{i}$, where $I_A$ is the indicator function of the event $A$. Thus, since $\nu_{f_i}$ is shift invariant, and since $Q_2$ is independent of the first $(N_i+1)p+1$ coordinates, we have  
\begin{align*}
    \E_{\nu_{f_i}}[C_n(\bY_0^{n-1})]&=(n-n_2)\Pro_{\mu}[A]=(n-n_2)\Pro_{\mu}\sparenv*{Q_2\cap\set*{\bY_{0}^{N_ip}\in S_i}}\\
    &=(n-n_2)\Pro_{\mu}\sparenv*{Q_2}\Pro_{\mu}\sparenv*{\bY_{0}^{N_ip}\in S_i}\\
    &\geq(n-n_2)\parenv*{\frac{1}{n_2}\parenv*{1-\frac{1}{N_i}}}(1-\varepsilon).
\end{align*}
Since $\delta_i$ is vanishing, for all sufficiently large $i$ (independent of $n$) we have $R_\varepsilon(X,Y,\mu)+\delta_i-1<0$. Thus, by combining the lower bound on $\E_{\nu_{f_i}}[C_n(\bY_0^{n-1})]$ with \eqref{eq:EqualityExp} we obtain
\begin{align*}
    \Pro_{\nu_{f_i}}[\bX_0\neq \bY_0]&=\E_{\nu_{f_i}}\sparenv*{\frac{1}{n}d\parenv*{\bX_0^{n-1},\bY_0^{n-1}}}\\
    &\leq \frac{n-n_2}{n}\cdot \frac{1}{n_2}\parenv*{1-\frac{1}{N_i}}(1-\varepsilon)(N_ip+1)(R_\varepsilon(X,Y,\mu)+\delta_i-1)+1.
\end{align*}
Taking the limit as $n\to \infty$ we get 
\begin{align*}
    \Pro_{\nu_{f_i}}[\bX_0\neq \bY_0]
    &\leq \frac{1}{(N_i+1)p+1}\parenv*{1-\frac{1}{N_i}}(1-\varepsilon)(N_ip+1)(R_\varepsilon(X,Y,\mu)+\delta_i-1)+1.
\end{align*}
We recall that $N_i\to \infty$ and $\delta_i\to 0$ as $i\to \infty$, and again by taking the limit as $i\to\infty$ we have
\begin{align*}
    \inf&\set*{\Pro_{\nu_g}[\bX_0\neq \bY_0]; g:Y\to X \text{ is a stationary coding function}} \\
    &\leq \lim_{i\to \infty}\Pro_{\nu_{f_i}}[\bX_0\neq \bY_0]\\
    &\leq\lim_{i\to \infty }\frac{1}{(N_i+1)p+1}\parenv*{1-\frac{1}{N_i}}(1-\varepsilon)(N_ip+1)(R_\varepsilon(X,Y,\mu)+\delta_i-1)+1 \\
    &= (1-\varepsilon)R_{\varepsilon}(X,Y,\mu)+\varepsilon.
\end{align*}
Taking $\varepsilon\to 0$, 
\[\inf\set*{\Pro_{\nu_g}[\bX_0\neq \bY_0]; g:Y\to X \text{ is a stationary coding function}}\leq R_0(X,Y,\mu). \]
\end{IEEEproof} 

We now have the first component of the proof of Theorem~\ref{th:SBCchar}. The second part of the proof requires the well known fact that stationary coding functions may be approximated by sliding-block-codes.

\begin{lemma}[{{\cite[Theorem 3.1]{gray1975sliding}}}]
\label{lem:SBCapprox}
Let $X,Y\subseteq \Sigma^\Z$ be shift spaces, $\mu\in M(Y)$, and $g:Y\to X $ be a stationary coding function. Then for any $\varepsilon>0$, there exists a sliding-block-code  function $\hat{f}:Y\to X$ such that 
\[\Pro_\mu\sparenv*{g(\bY)\neq \hat{f}(\bY)}<\varepsilon.\]
\end{lemma}

We are now ready to prove Theorem~\ref{th:SBCchar}.

\begin{IEEEproof}[Proof of Theorem~\ref{th:SBCchar}]
Fix $\varepsilon>0$. By Theorem~\ref{th:JoinTight}, there exists a stationary coding function $g:Y\to X$ such that
\[\Pro_{\mu}[g(\bY)_0\neq \bY_0]=\Pro_{\nu_{g}}[\bX_0\neq \bY_0]< R_0(X,Y,\mu)+ \frac{1}{2}\varepsilon.\] By Lemma~\ref{lem:SBCapprox},  there exists a sliding-block-code function $\hat{f}:Y\to X$ and a set $E$ with $\mu(E)>1-\frac{1}{2}\varepsilon$ such that $g$ coincides with $\hat{f}$ on $E$. We now have,
\begin{align*}
    \Pro_{\nu_{f}}[\bX_0\neq \bY_0]&=\Pro_{\mu}[\bY_0\neq \hat{f}(\bY)_0]\\
    &=\Pro_{\mu}[\set{\bY_0\neq \hat{f}(\bY)_0}\cap E]+\Pro_{\mu}[\set{\bY_0\neq \hat{f}(\bY)_0}\cap E^C]\\
    &\leq \Pro_{\mu}[g(\bY)_0\neq \bY_0]+\Pro_{\mu}[E^C]< R_0(X,Y,\mu)+\frac{1}{2}\varepsilon+\frac{1}{2}\varepsilon=R_0(X,Y,\mu)+\varepsilon.
\end{align*}
The upper bound $R_0(X,Y,\mu)\leq  \Pro_{\nu_{f}}[\bX_0\neq \bY_0] $ is immediate by Proposition~\ref{prop:JoinBound}.
\end{IEEEproof}


\section{Conclusion}\label{sec:Disc}

In this work, we introduced the Quantized-Constraint Concatenation (QCC) scheme, providing a new general framework for implementing error-correction in constrained systems. We have shown that by embedding codewords of an error-correcting code in a constrained system by way of quantization, it is possible to correct $\Theta(n)$ errors (with respect to the code's length $n$). We discovered that the asymptotic error-correction capabilities of our method for a given constrained system are determined by a new fundamental parameter of the constrained system -- its covering radius -- which bounds the amount of noise caused by the quantization process. Unlike previous methods, such as concatenation and reverse concatenation, the embedding into the constrained system is not reversible, hence the term quantization.

We presented two different definitions for the covering radius of a constrained system, showing it as a combinatorial and as a probabilistic notion. While the combinatorial definition takes into account the worst-case scenario (deep holes), in the probabilistic approach, the essential covering radius ignores rare cases and therefore allows a smaller covering radius. We have studied the properties of the essential and combinatorial covering radii, and provided general lower and upper bounds. 

While the covering radius of constrained systems is of independent intellectual merit, let us put our results in the context of the QCC scheme. Consider $X_{0,k}$, the $(0,k)$-RLL system described in Example~\ref{ex:0kRLL}. Using the coding scheme presented in \cite[Theorem 1]{gabrys2020segmented}, it is possible to correct up to $O(\sqrt{n})$ errors. However, using QCC with the combinatorial covering radius (which in that case is $\frac{1}{k+1}$), since there exist error-correcting codes with non-vanishing rate capable of correcting up to $(\frac{1}{4}-\delta)n$ errors (for every $\delta>0$), we obtain codes with non-vanishing rate capable of correcting up to $ (\frac{1}{4} -\frac{1}{k+1} -\delta)n$ channel errors. On the other hand, we may use the essential covering radius of $X_{0,k}$ to bound the \emph{probable} quantization noise. In that case, since
\[R_{\frac{1}{2}}(X_{0,k},[2]^\Z,\muni)\leq  R_0(X_{0,k},[2]^\Z,\muni)=\frac{1}{2(2^{k+1}-1)},\]
using the QCC, it is possible to find error-correcting codes such that by removing at most half of the codewords (which asymptotically does not effect the rate), it is possible to improve our error correction capability to $(\frac{1}{4} -\frac{1}{2(2^{k+1}-1)} -\delta)n$ channel errors. Previous lower bounds on the possible rates for error-correcting constrained codes have been  established in previous work \cite{marcus1992improved,kolesnik1991generating}, via somewhat non-constructive methods. A certain advantage of our scheme is its simplicity and constructive nature. The advantage of our scheme is its simplicity and constructive nature.   

\bibliographystyle{IEEEtranS}
\bibliography{Biblio_Commented}

\end{document}